\documentclass[chicago]{emulateapj-rtx4}

\usepackage{amsmath}
\usepackage{amssymb}
\usepackage{graphicx}
\usepackage{wasysym}
\usepackage{multirow}
\usepackage{mdframed}
\usepackage{lscape}

\usepackage[breaklinks,colorlinks,citecolor=blue]{hyperref}

\newcommand{\methanol}{\mbox{CH$_3$OH}}
\newcommand{\twelveco}{\mbox{$^{12}$CO}}
\newcommand{\twelvecotto}{\mbox{$^{12}$CO\,2--1}}
\newcommand{\thirteenco}{\mbox{$^{13}$CO}}
\newcommand{\ceighteeno}{\mbox{C$^{18}$O}}
\newcommand{\fmh}{\mbox{H$_2$CO}}
\newcommand{\water}{\mbox{H$_2$O}}
\newcommand{\amm}{\mbox{NH$_3$}}
\newcommand{\ammone}{\mbox{\amm{}\,(1,1)}}
\newcommand{\ammtwo}{\mbox{\amm{}\,(2,2)}}
\newcommand{\ammthree}{\mbox{\amm{}\,(3,3)}}
\newcommand{\ntwohplus}{\mbox{N$_2$H$^+$}}
\newcommand{\hthirteencoplus}{\mbox{H$^{13}$CO$^+$}}
\newcommand{\kms}{\mbox{km\,s$^{-1}$}}

\newcommand{\cd}{\mbox{cm$^{-2}$}}
\newcommand{\cc}{\mbox{cm$^{-3}$}}
\newcommand{\lsol}{\mbox{L$_\odot$}}
\newcommand{\msol}{\mbox{M$_\odot$}}

\slugcomment{Not to appear in Nonlearned J., 45.}

\shorttitle{Initial Fragmentation in the IRDC G28.53$-$0.25}
\shortauthors{Lu et al.}

\begin{document}

\title{Initial Fragmentation in the Infrared Dark Cloud G28.53$-$0.25}
	
\author{Xing Lu\altaffilmark{1,2,3,4}} \author{Qizhou Zhang\altaffilmark{2}} \author{Ke Wang\altaffilmark{5}} \author{Qiusheng Gu\altaffilmark{1,3,4}}
\affil{$^1$ School of Astronomy and Space Science, Nanjing University, Nanjing, Jiangsu 210093, China; \href{mailto:xlu@cfa.harvard.edu}{xlu@cfa.harvard.edu}}
\affil{$^2$ Harvard-Smithsonian Center for Astrophysics, 60 Garden Street, Cambridge, MA 02138}
\affil{$^3$ Key Laboratory of Modern Astronomy and Astrophysics (Nanjing University), Ministry of Education, Nanjing 210093, China}
\affil{$^4$ Collaborative Innovation Center of Modern Astronomy and Space Exploration, Nanjing 210093, China}
\affil{$^5$ European Southern Observatory, Karl-Schwarzschild-Str.\ 2, D-85748 Garching bei M\"{u}nchen, Germany}

\begin{abstract}
To study the fragmentation and gravitational collapse of dense cores in infrared dark clouds (IRDCs), we have obtained submillimeter continuum and spectral line data as well as multiple inversion transitions of \amm{} and \water{} maser data of four massive clumps in an IRDC G28.53$-$0.25. Combining single dish and interferometer \amm{} data, we derive the rotation temperature of G28.53. We identity 12 dense cores at 0.1~pc scale based on submillimeter continuum, and obtain their physical properties using \amm{} and continuum data. By comparing the Jeans masses of cores with the core masses, we find that turbulent pressure is important in supporting the gas when 1~pc scale clumps fragment into 0.1~pc scale cores. All cores have a virial parameter smaller than 1 assuming a inverse squared radial density profile, suggesting they are gravitationally bound, and the three most promising star forming cores have a virial parameter smaller than 1 even taking magnetic field into account. We also associate the cores with star formation activities revealed by outflows, masers, or infrared sources. Unlike what previous studies suggested, MM1 turns out to harbor a few star forming cores and is likely a progenitor of high-mass star cluster. MM5 is intermediate while MM7/8 are quiescent in terms of star formation, but they also harbor gravitationally bound dense cores and have the potential of forming stars as in MM1.
\end{abstract}

\keywords{stars: formation $-$ ISM: molecules}

\section{INTRODUCTION}\label{sec:intro}

Massive infrared dark clouds (IRDCs) in the Galaxy have been recognized as the birthplace of high-mass ($M>8~\textrm{M}_\odot$) stars, since they are massive, cold, and dense \citep{simon2006,pillai2006,rathborne2006}. Millimeter/sub-millimeter interferometric observations toward IRDCs have found 0.1~pc scale fragments or cores that might harbor high-mass stars \citep{zhang2009,zhang2011,wang2011,wang2014,beuther2013,peretto2013}. These cores are often associated with molecular outflows, but lack complex organic molecular line emission that are tracing hot cores. Therefore, they are the most promising specimen for studying the very early phase of high-mass star formation. 

One important question concerning these cores in IRDCs is how they fragment and evolve, until high-mass stars are born. \citet{zhang2009} found that in IRDCs parsec-scale clumps fragment into 0.1~pc scale cores, and \citet{wang2011,wang2014} found that 0.1~pc scale cores further fragment into $<$$0.1$~pc scale condensations, all of which are more massive than the corresponding thermal Jeans masses but are consistent with turbulent Jeans masses. Therefore, in IRDCs turbulent pressure is essential in supporting the fragmentation, so that massive cores are able to form and grow. \citet{pillai2011} studied 0.1~pc scale cores in two high-mass star forming regions, and found that all the cores are gravitationally bound, with their gravitational mass far exceeding the virial mass, therefore turbulence is not sufficient to stop cores from fast collapsing. 

To study the impact of turbulence in the initial fragmentation and core growth in IRDCs, we select four clumps in an IRDC G28.53$-$0.25 (G28.53 hereafter). G28.53 has a kinematic distance of 5.4~kpc \citep{rathborne2006} and a luminosity of $\sim$3500~\lsol{} \citep{rathborne2010}. It has been mapped in 1.2 mm continuum with the IRAM 30m single-dish telescope \citep[][also see \autoref{fig:iram}]{rathborne2006,rathborne2010}, which reveals a total mass of $\sim$$10^4$~\msol{}. Ten continuum peaks are identified, each of which is a few hundreds of \msol{} and of $\sim$1~pc scales, therefore they are typical clumps that form massive stars. \citet{rathborne2010} also classified these clumps as active, intermediate, or quiescent in terms of star formation based on the infrared emission. Among them, the most massive ($\sim$$10^3$~\msol{}) one at the center of G28.53, MM1, is classified as quiescent. A less massive but more luminous clump at the southern end of the cloud, MM5, is classified as intermediate. Two clumps in the southeast corner, MM7 and MM8, are classified as quiescent. \citet{sanhueza2012} obtained 3~mm spectral lines of the ten clumps and confirmed the classification of \citet{rathborne2010} based on chemistry.

Interferometric observations at angular resolutions of 1\arcsec{}--2\arcsec{} toward MM1 revealed further fragmentation \citep{rathborne2008,swift2009}. \citet{rathborne2008} resolved a 0.1~pc scale core in MM1 into three smaller condensations. Although these condensations do not present detectable spectral line emission at an arcsecond resolution, the single-dish observations of MM1 suggested active star formation given broad line wings and detection of SiO emission which is a typical shock tracer. \citet{swift2009} mapped MM1 with the Submillimeter Array (SMA) in 345~GHz and found a massive ($\sim$60~\msol{}) core known as `core~2', which is different with the one found by \citet{rathborne2008}. No CO outflows or hot core tracers were found in this core. 

However, given its large amount of gas and its location at the center of the gravitational potential well of G28.53, it is puzzling that MM1 is more quiescent than MM5. Previous single-dish observations might miss the deeply embedded cores in MM1 that are actively forming stars. The two interferometric studies detected a few dust cores but lacked molecular lines to trace the star formation, and the CO emission at 345~GHz as an outflow tracer was filtered by the interferometer. Therefore, it is worth revisiting the 0.1~pc scale fragmentation within MM1, and compare it with the `intermediate' MM5 and `quiescent' MM7 and MM8.

In this paper, we will use high angular resolution \amm{} line observations to trace kinematics and determine gas temperatures. \amm{} itself is also useful in tracing dense gas with a critical density of $\sim$$10^4$~\cc{} \citep{swade1989}. We will also use millimeter continuum and spectral lines to resolve 0.1~pc scale cores in the four clumps and trace star formation. The combination of kinematics (e.g.\ linewidth, velocity components along the line of sight) and temperatures from \amm{} lines and dust emission from millimeter continuum will allow us to determine the properties of cores reliably. We will introduce details of the observations in Section~\ref{sec:obs}. Then we present the results in Section~\ref{sec:results}, and discuss the fragmentation and star formation in the clumps in Section~\ref{sec:discussions}. In Section~\ref{sec:conclusions} we summarize the main results and present our conclusions.

\addtocounter{footnote}{5}

\section{OBSERVATIONS}\label{sec:obs}

\subsection{SMA observations}\label{subsec:smaobs}

\subsubsection{230~GHz observations of G28.53 MM5, MM7, \& MM8}\label{subsubsec:smaobs_mm578}

In 2010, we observed G28.53 MM5, MM7, and MM8 with seven antennas of the Submillimeter Array\footnote{The Submillimeter Array is a joint project between the Smithsonian Astrophysical Observatory and the Academia Sinica Institute of Astronomy and Astrophysics and is funded by the Smithsonian Institution and the Academia Sinina.} \citep[SMA;][]{ho2004} in the compact configuration at 230~GHz band. Frequency-dependent bandpass solutions were obtained by observing quasars 3C454.3. Time-dependent gain solutions were obtained by observing the quasar 1911$-$201 every 20 minutes. The flux calibration was done using Titan and Neptune. The correlator covers rest frequencies of 216.7--220.7~GHz in the lower side band, and 228.7--232.7~GHz in the upper sideband, with a uniform channel width of 0.812~MHz, which is equivalent to 1.1~\kms{} at 230~GHz. System temperatures were 100--120~K and opacity at 225~GHz was 0.06--0.1. The observations are summarized in \autoref{tab:obs}. 

The visibility data were calibrated using \mbox{MIR}\footnote{\url{https://www.cfa.harvard.edu/~cqi/mircook.html}}. Calibrated data were then inspected and imaged using \mbox{MIRIAD}\footnote{\url{http://www.cfa.harvard.edu/sma/miriad/}} \citep{sault1995} and \mbox{CASA}\footnote{\url{http://casa.nrao.edu/}} \citep{mcmullin2007}. Continuum emission was extracted by averaging line free channels in the visibility domain, then was imaged using combined data from both sidebands. The spectral lines were cut out from the continuum-subtracted visibility data and were imaged separately. All images were CLEANed with a robust parameter of 0.5 to obtain a balance between good sensitivity and side lobe suppression. Image properties are listed in \autoref{tab:image}.

\subsubsection{Archival 230~GHz data of G28.53 MM1}\label{subsubsec:smaobs_mm1}

We used the SMA 230~GHz archival data of G28.53 MM1. The observations were carried out in the 230~GHz band in two tracks in 2009. The pointing center is at ($18^h44^m18.0^s, -3\arcdeg59\arcmin23.0\arcsec$). The SMA was in the compact configuration with seven antennas in the array. Frequency-dependent bandpass solutions were obtained by observing the quasar 3C273. Time-dependent gain solutions were obtained by observing the quasar 1751+096 every 28.5 minutes. The flux calibration was done using Callisto. The two tracks both cover 2~GHz in each sideband of the SMA, with a uniform channel width of 0.812~MHz, which is equivalent to 1.1~\kms{} at 230~GHz. The tracking frequencies of the two tracks are different, so that $J=2\text{--}1$ lines of CO isotopologues and the \methanol{} line at 229.759~GHz were observed in one track and the \fmh{} line at 225.698~GHz was observed in the other. System temperatures were $\sim$80--160~K and opacity at 225~GHz was $\sim$0.06--0.08. The observations are summarized in \autoref{tab:obs}. The visibility data were calibrated and imaged in the same manner as in the previous section. Image properties are listed in \autoref{tab:image}.

\subsection{VLA and GBT \amm{} observations}\label{subsec:nh3obs}

We observed five positions in G28.53 using the National Radio Astronomy Observatory (NRAO)\footnote{The National Radio Astronomy Observatory is a facility of the National Science Foundation operated under cooperative agreement by Associated Universities, Inc.} Very Large Array (VLA) in the D configuration in two observation runs in 2010 January, to obtain the \amm{}\,$(J,K)=(1,1)$ and (2,2) transitions. The 3.125~MHz bands were each split into 128 channels with a channel spacing of 24.4~kHz, equivalent to 0.3~\kms{}. The two bands simultaneously covered the main and two inner satellite transitions of the \ammone{} line, as well as the main and one inner satellite transitions of the \ammtwo{} line. Coordinates of these positions are listed in \autoref{tab:obs}.

The quasars 0137+331 (3C48) and 1331+305 (3C286) were used for flux calibration. Bandpass calibration was done with observations of 1229+020 (3C273) and 2253+161 (3C454.3). Gain calibration was performed with periodical observations of 1851+005 every 15 minutes. The visibility data were calibrated using CASA.

The NRAO Green Bank Telescope (GBT) was used to observe a $8\arcmin\times7.5\arcmin$ region in G28.53 in 2010 February, to obtain the \amm{}\,$(J,K)=(1,1)$ and (2,2) transitions. The data were calibrated and imaged using \mbox{GBTIDL}\footnote{\url{http://gbtidl.nrao.edu/}}. The \amm{} images were then Fourier transformed into the visibility domain based on the VLA visibility models.

Then the VLA and GBT visibility data were combined and imaged with \mbox{MIRIAD}. The combined images keep the native channel width of the VLA data. Image properties are listed in \autoref{tab:image}.

We also observed the \ammthree{} transition using the VLA with the WIDAR correlator in its D configuration towards G28.53 MM1 in 2010 May. The calibrators are listed in \autoref{tab:obs}. The data were calibrated and imaged using \mbox{CASA}. Image properties are listed in \autoref{tab:image}.

\subsection{VLA maser observations}\label{subsec:maserobs}

The \water{} maser at 22~GHz and the class I \methanol{} maser at 25~GHz were observed using the VLA with the WIDAR correlator in 2010 November, toward two positions, MA1 and MA5 which are close to MM1 and MM5. Details of observations are listed in \autoref{tab:obs}. The data were calibrated and imaged using \mbox{CASA}. Image properties are listed in \autoref{tab:image}.

\section{RESULTS}\label{sec:results}

\subsection{\amm{} emission and temperature}\label{subsec:results_nh3}
We detected \ammone{} and (2,2) emission with both VLA and GBT\@. In the combined images, we found \ammone{} and (2,2) emission associated with from MM1 to MM8, enabling us to derive temperatures of these clumps. The rotation temperature derived from these two transitions increases almost linearly with the kinetic temperature and differs by $<$3~K up to 20~K \citep{walmsley1983}, therefore can be used an equivalence of the kinetic temperature.

We simultaneously fitted the \ammone{} and (2,2) spectra from the combined images, to derive the best fitted linewidth, line intensity, and optical depth at the same time. The fitting model includes three gaussians for the (1,1) main component and inner satellite components, and one gaussian for the (2,2) main component, which are all in the form of
\begin{equation}
I(v)=I_0\left\{1-\exp\left[-\tau_0\exp\left(-\frac{1}{2}\left(\frac{v-v_0}{\sigma_v}\right)^2\right)\right]\right\}
\end{equation} 
in which $I_0$ is the effective line intensity, $\tau_0$ is the optical depth, $v_0$ is the central velocity, and $\sigma_v$ is the velocity dispersion. The central velocity was determined using a cross-correlation between the spectra and a model spectrum and was then fixed during the fitting. We put constraints so that the ratio of optical depths of the main and satellite lines of \ammone{} is 1/0.278 for both satellite lines, all four gaussians have the same velocity dispersion, and the velocity separations between the main and two inner satellite lines are 7.47 and 7.57~\kms{} \citep{mangum1992}, then we minimized the difference between the spectra and the model, using the Levenberg-Marquardt algorithm implemented by the lmfit Python package\footnote{\url{https://github.com/newville/lmfit-py}}. The rotation temperature and \amm{} column density were then derived from the best-fitted parameters as in \citet{Ho1983} and \citet{mangum1992}. The temperature map is shown in \autoref{fig:temperature}a. The rotation temperature throughout G28.53 is $\lesssim$17~K. Even around the dust core `core~2' and the masers, it is not showing increased temperature. 

The \amm{} spectra within MM1 present two velocity components, at $\sim$85~\kms{} and $\sim$87~\kms{} respectively. Similar multiple components have been detected in the central regions of a few filamentary IRDCs \citep{zhang2011,henshaw2013}. For these spectra, the general fitting procedure above derived overall properties which blended the two components and derived a temperature that is usually a few~K different from the true temperature of either component (cf.\ \autoref{fig:temperature}a and \autoref{fig:2comp}). 

We attempted to fit the two velocity components simultaneously and derive a temperature map for each component. However, it was unsuccessful thanks to the limited signal-to-noise ratios in the spectra. Thus, we averaged the spectra within the cores in MM1 defined in Section~\ref{subsec:results_fragmentation} to increase the signal-to-noise ratio, and fitted them with a model containing two components each of which includes four gaussians as in the single velocity component case. The two central velocities were both fixed during the fitting. The fitting results are shown in \autoref{fig:2comp} and \autoref{tab:2comp}. Typical temperatures are 13--17~K, while the temperature of the 85~\kms{} component is in general 2~K higher that of the 87~\kms{} component, suggesting that there are indeed two distinct components. We obtained \amm{} column densities of the two velocity components, which were used to determine the ratio of dust masses in the two components in Section~\ref{subsec:results_fragmentation}.

We also derived the rotation temperature using the VLA \amm{} data only. Without GBT data, the interferometer filters out extended emission and has more contribution from compact structures. However, the change of the temperature map is insignificant, as shown in \autoref{fig:temperature}b. The highest temperature is $\gtrsim$20~K which is found in MM1, while typical temperatures are still 13--17~K as in \autoref{fig:temperature}a. Note that the two velocity components also show up in MM1 in the VLA data. We fitted the average spectra of MM1-p1 with two velocity components and showed the result in \autoref{fig:vlaonlytemperature}. The temperature of the 87~\kms{} component is consistent with that derived from the combined data, while the temperature of the 85~\kms{} component is up to $\sim$20~K.

In addition, we detected \ammthree{} emission in MM1 with VLA\@. Unlike \ammone{} and (2,2), the (3,3) emission is concentrated. We marked the three positions exhibiting strong (3,3) emission in \autoref{fig:temperature}b as T1--T3, as well as `core~2' which presents weak ($5\sigma$) (3,3) emission, with their spectra shown in insets. T1 presents two velocity components, at $\sim$85 and $\sim$87~\kms{}. The other three spectra all center at $\sim$87~\kms{}. We fitted a single gaussian to the spectra of `core~2', T2, and T3, and two gaussians at 85.5 and 87.9~\kms{} to the spectrum of T1. The FWHM linewidth of `core~2' is up to 4.0~\kms{}, while those of T2 and T3 are 0.5 and 1.2~\kms{}, respectively. The FWHM linewidths of the two components of T1 are 2.6 and 1.8~\kms{} respectively. It is worth noting that T2 and T3 are next to the \water{} maser W4, whose positions are consistent with the possible CO outflow lobes surrounding W4 as discussed in Section \ref{subsec:results_masers}. These two \ammthree{} emission peaks might suggest outflow heating and turbulence injection given the high excitation temperature of \ammthree{} (123.5~K) and the large linewidths we detected. This is similar to what \citet{wang2012} found in the IRDC G28.34.

\subsection{Fragmentation in clumps}\label{subsec:results_fragmentation}

The SMA observations of the four clumps revealed 0.1~pc scale fragmentation. All peaks with fluxes larger than 5$\sigma$ in the SMA 230~GHz dust emission images were identified. These $\sim$0.1~pc scale peaks are in agreement with the definition of cores \citep[e.g.][]{zhang2009}. Then we fitted 2D gaussian functions to them and obtained their positions and deconvolved sizes. We also applied the primary beam correction to the images and obtained the corrected fluxes of the cores. The results are listed in \autoref{tab:cores}.

We assumed a gas-to-dust mass ratio of 100, and thermal equilibrium between dust and gas, then the mass is \citep{beuther2005}:
\begin{equation}\label{eq:mcore}
M_\mathrm{core}=\frac{2.0\times10^{-2}}{J_\nu(T_\mathrm{rot})}\frac{F_\nu}{\mathrm{Jy}}\left(\frac{D}{\mathrm{kpc}}\right)^2\left(\frac{\nu}{\mathrm{1.2\ THz}}\right)^{-3-\beta}\ \mathrm{M_\odot},
\end{equation}
in which $J_\nu(T_\mathrm{rot})=\left[\exp\left(h\nu/kT_\mathrm{rot}\right)-1\right]^{-1}$, and the dust emissivity index $\beta=1.5$ for MM1 and MM8, 1.7 for MM5, and 1.0 for MM7 \citep{rathborne2010}. 

We also reported errors of $M_\mathrm{core}$ in \autoref{tab:cores}, assuming that $M_\mathrm{core}$ is only dependent on $T_\mathrm{rot}$, $F_\nu$, and $D$. The error in the kinematic distance was assumed to be 10\% \citep{reid2009}. The errors in the fluxes and temperatures listed in the table are from gaussian fittings without accounting for observational uncertainties. The observational uncertainty of the fluxes are mainly due to flux calibration as described in Section \ref{subsec:smaobs}, which is typically 20\%, while that of the temperatures depends on the signal-to-noise ratio of the \amm{} lines \citep[e.g.][]{li2003}, which is typically $<$0.5~K for our spectral data with a signal-to-noise ratio of $>$20. Accumulating all the errors listed above, typical errors in $M_\mathrm{core}$ are 30\%, which should be interpreted as lower limits given the unknown uncertainty in $\beta$, possible discrepancy in the temperatures of dust and \amm{}, and the uncertainty in gas-to-dust mass ratio.

For MM1, using the compact configuration data, we resolved it into six cores at an angular resolution of $\sim$2.3\arcsec, labelled as MM1-p1 to MM1-p6 in \autoref{fig:sma_MMs}a. We detected \methanol{} line emission at 87~\kms{} towards MM1-p1, which is the only core that this line is detected among all cores we identified in the four clumps.

The cores in MM1 all exhibit two velocity components in \amm{} lines. \amm{} emission of both components morphologically matches the dust emission. We considered the component at $\sim$87~\kms{}, because both the \methanol{} line emission we detected and the single-dish \ntwohplus{} and \hthirteencoplus{} lines \citep{rathborne2008} are at $\sim$87~\kms{}. After deriving masses using the parameters listed in \autoref{tab:cores}, we determined the ratios of masses in the two components based on the \amm{} column densities in \autoref{tab:2comp}, and scaled the masses of the cores.

Among the six cores in MM1, MM1-p1 is consistent with `core~2' reported in \citet{swift2009} while MM1-p2 is consistent with the core reported in \citet{rathborne2008}, both using 345~GHz dust continuum emission. \citet{rathborne2008} resolved MM1-p2 into three condensations, at an angular resolution of $\sim$2\arcsec. The three condensations have sizes of $\sim$0.06~pc and masses of 9--20~\msol{}. In addition, in the dust emission map of \citet{swift2009}, we found weak features at 3--5$\sigma$ levels that are coincident with the other four cores in MM1. 

For MM5, we identified three cores, which are labelled as MM5-p1 to MM5-p3 in \autoref{fig:sma_MMs}b. MM7 and MM8 are at the southeast corner of G28.53, in a filamentary structure isolated from the main part of the cloud. We resolved MM7 to one core, and MM8 to two cores, labelled in \autoref{fig:sma_MMs}c. The properties of these cores are listed in \autoref{tab:cores}.

Aside from dust continuum emission, we detected a few spectral lines toward these cores using the SMA\@. As shown in \autoref{fig:sma_spec}, MM1-p1 exhibits \twelveco{}, \thirteenco{}, and \ceighteeno{}\,2--1 lines, as well as \methanol{}\,8($-$1,8)--7(0,7) and \fmh{}\,3(1,2)--2(1,1) lines. MM5-p1 exhibits \twelvecotto{} and \fmh{}\,3(0,3)--2(0,2) lines, but little \thirteenco{}, \ceighteeno{} or \methanol{} line emission. MM7-p1 and MM8-p1 exhibit only \twelveco{} line. \methanol{} and \fmh{} lines are tracers of dense gas given their large critical densities (10$^{5\text{--}6}$~\cc{}), while \twelveco{} line could trace outflows. The relationship between these lines and star formation in the cores will be discussed in Section \ref{subsec:disc_mm15}.

\subsection{Protostellar outflows}\label{subsec:results_outflows}
The SMA observations covered the \twelvecotto{} line, which can trace molecular outflows driven by protostars. We detected CO outflows in MM1 and MM5. MM7 and MM8 also present CO emission, but do not show any signature of outflows.

The CO channel maps in \autoref{fig:CO_MM1} reveal two parallel bipolar outflows, associated with MM1-p1 and MM1-p2. The CO emission associated with MM1-p1 presents two blue shifted components at 81 and 83~\kms{} with respect to the system velocity of 87~\kms{}, on opposite sides of MM1-p1. At more redshifted velocities between 90--100~\kms{}, CO emission is also found around MM1-p1. It is consistent with a bipolar outflow nearly parallel to the plane of the sky, with the two components at 81 and 83~\kms{} tracing its near side while the emission at more redshifted velocities tracing its far side \citep[e.g.\ Figure~8 of][]{wu2009}. Similarly, the CO emission associated with MM1-p2 reveals two components at 90--92~\kms{} and 92--94~\kms{} on two opposite sides of the core, which might trace the red-shifted component of a bipolar outflow in the plane of the sky. We found a CO emission gap at 85--88~\kms{}, indicating the SMA filtered out the extended CO emission around the systematic velocity.

A \water{} maser detected in \citet{wang2006} (W1, Section~\ref{subsec:results_masers}) is in the projected path of the CO outflow associated with MM1-p2. Its velocity is 85~\kms{}, while the CO velocity here is 94--96~\kms{} which is 10~\kms{} apart, therefore it is unlikely excited by the outflow. We also found compact \fmh{} line emission, which has a critical density of $\sim$$10^5$~\cc \citep{guzman2011}, at the same position of W1 but at the velocity of 89~\kms{}. It could be a dense core which is not detected by the SMA dust observations because of insufficient sensitivity at the edge of the primary beam.

We also found a bipolar outflow in MM5, shown in the CO channel maps in \autoref{fig:CO_MM5}. The outflow is oriented in a north-south direction, with a blue shifted component at 79--83~\kms{} and a red shifted component at 90--96~\kms{}. The central velocity, $\sim$87~\kms{}, is consistent with the velocities of the cores. However, the three cores in MM5 are closely packed, thus we can not identify which one is driving this outflow.

Assuming LTE and optically thin \twelveco{} emission in the outflow wings, we derived outflow properties as in \citet{wang2011}, listed in \autoref{tab:outflows}. The relative abundance [CO]/[H$_2$] was assumed to be 10$^{-4}$ \citep{blake1987}. We did not correct for inclination angles of the outflows. 

The dynamic ages of $\sim$10$^4$ years suggest very early evolutionary phases of these cores \citep[cf.][]{zhang2001,zhang2005,beuther2002}. Among the three outflows, the two in MM1 are more massive and more energetic than the one in MM5. All three outflows are comparable with the high-mass outflows in other IRDCs \citep[e.g.][]{wang2011, wang2014}.

\subsection{\water{} and \methanol{} masers}\label{subsec:results_masers}

The VLA observations of class I \methanol{} maser at 25~GHz did not detect any sources. The 22~GHz VLA observations revealed three \water{} masers around MM1, and none in MM5. Including the \water{} maser found by \citet{wang2006}, there are 4 \water{} masers around MM1, labelled as W1--W4 in Figures~\ref{fig:temperature} and \ref{fig:sma_MMs}a, among which W1 and W4 are within MM1. \autoref{tab:masers} summarizes properties of the masers.

\begin{figure}
\centering
\includegraphics[width=0.5\textwidth]{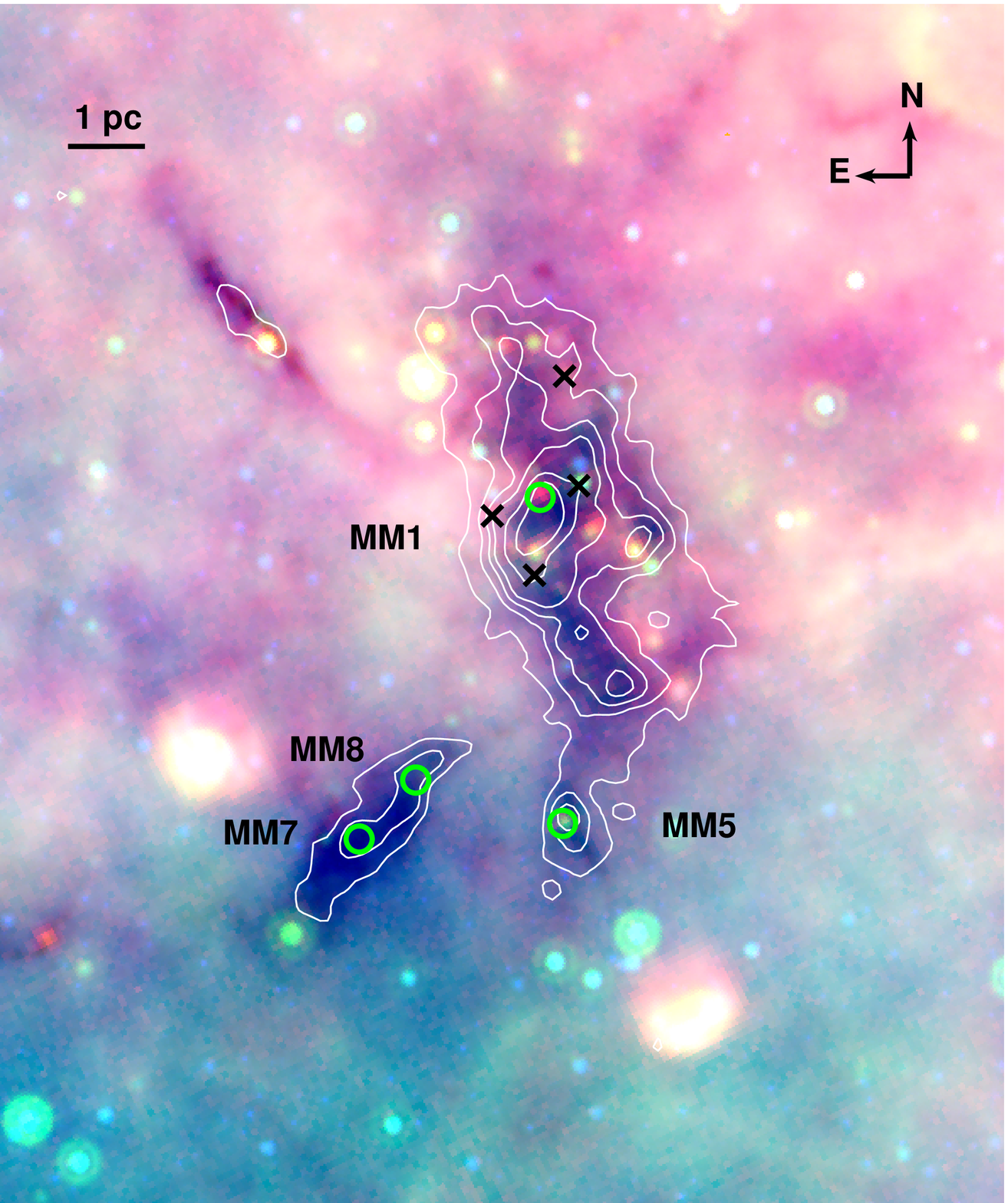}
\caption{\textit{Spitzer} and \textit{Herschel} composite image overlaid with 1.2~mm continuum emission of G28.53 taken from \citet{rathborne2006}. The colors represent emission at 8.0~\micron{} (blue), 24~\micron{} (green), and 70~\micron{} (red) \citep{churchwell2009,carey2009,molinari2010}. The contours levels are 30 (3$\sigma$), 60, 90, 120, 180, 240 mJy per beam. The four clumps, MM1, MM5, MM7, and MM8 are marked. The green circles mark MM1-p1, MM5-1, MM7-p1, and MM8-p1, the most massive cores in each clump. The four crosses mark the \water{} masers.}
\label{fig:iram}
\end{figure}

\begin{figure}
\begin{tabular}{p{8.5cm}p{8.5cm}}
\includegraphics[width=0.48\textwidth]{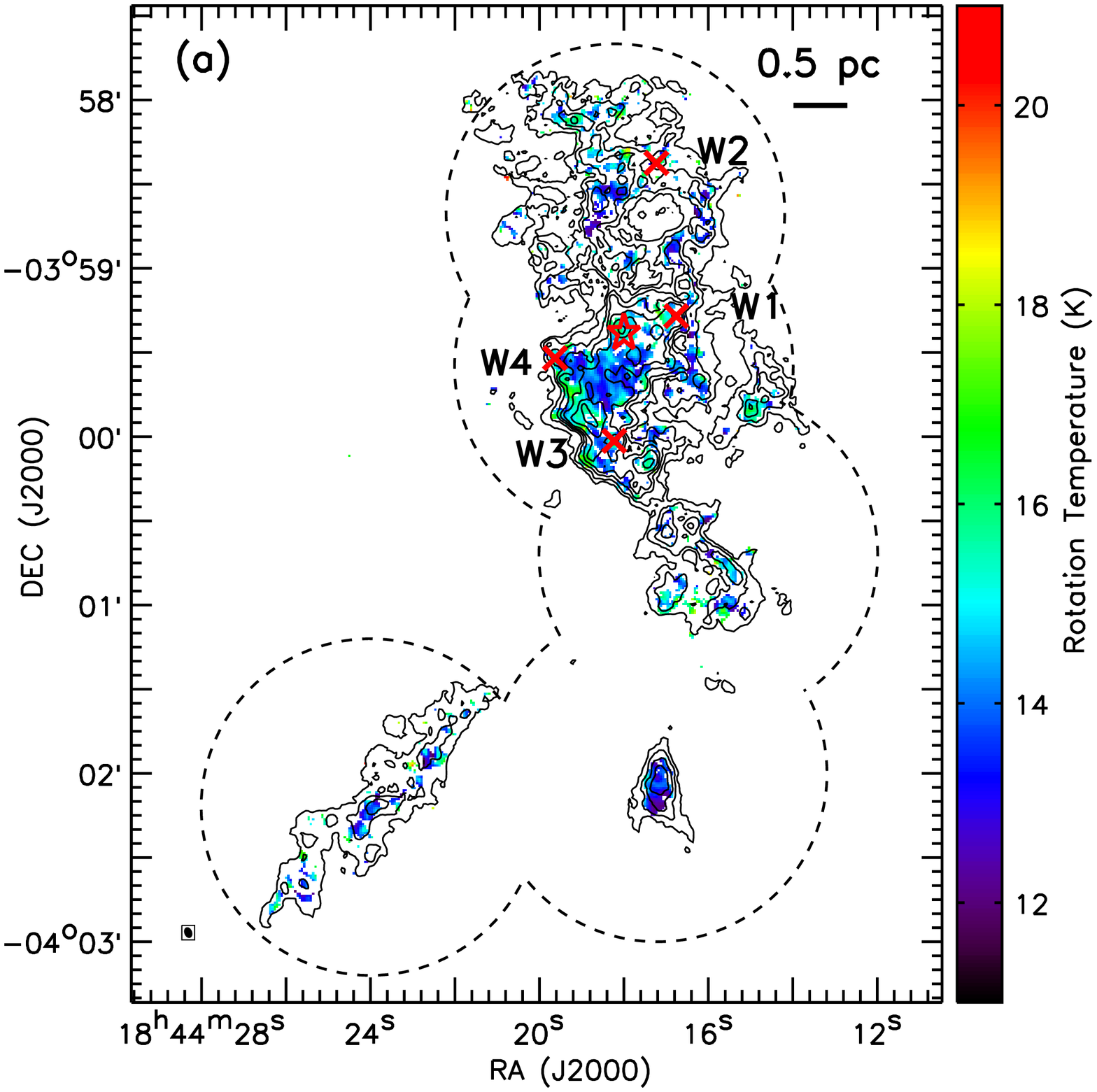} & \includegraphics[width=0.48\textwidth]{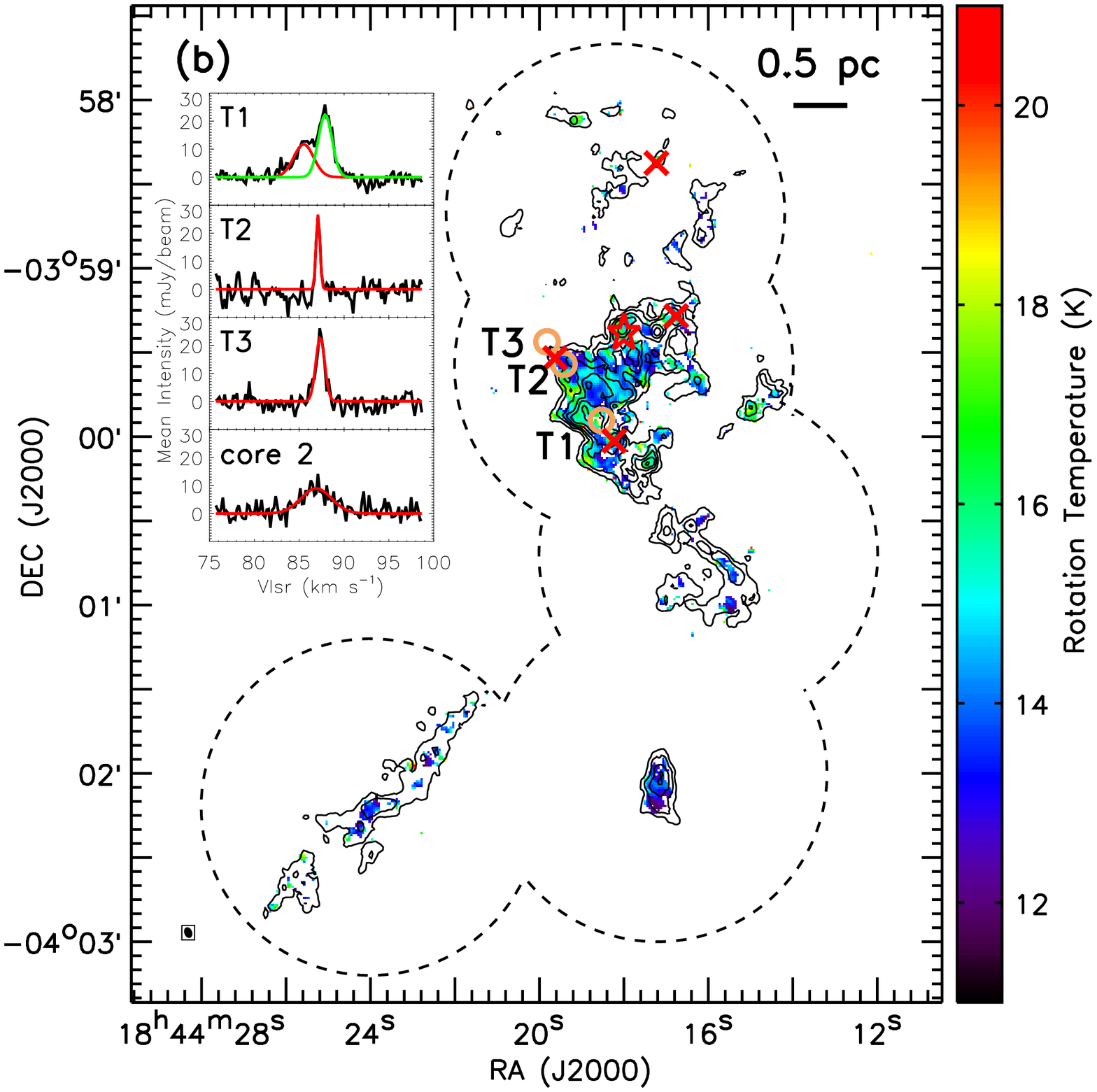}
\end{tabular}
\caption{\amm{} rotation temperature of G28.53. (a) Temperature map using the combined \mbox{GBT} and \mbox{VLA} data, overlaid with contours of the integrated intensities of the \ammone{} main hyperfine line. The contours are in steps of 5~\mbox{mJy\,beam$^{-1}$\,\kms{}}$\times$[3, 6, 9, 12, 15, 20, 25]. The dash circles represent the 2\arcmin{} primary beams of the \mbox{VLA} observations. `Core 2' in MM1 is marked with a star. The four \water{} masers are marked with crosses and labelled as W1--W4. (b) Temperature map using the \mbox{VLA} data only, overlaid with contours of the integrated intensities of the \ammone{} main hyperfine line. The contours are in steps of 5~\mbox{mJy\,beam$^{-1}$\,\kms{}}$\times$[3, 6, 9, 12, 15, 20]. The three positions exhibiting \ammthree{} emission are marked with circles and labelled as T1--T3, with their spectra and the best gaussian fitting shown in the inset. Note that the two maps are plotted using the same color scale from 11~K to 21~K to show their similarity, but the highest temperature in (a) is 19~K.}
\label{fig:temperature}
\end{figure}

\begin{figure}
\begin{tabular}{p{8.5cm}p{8.5cm}}
\includegraphics[width=0.5\textwidth]{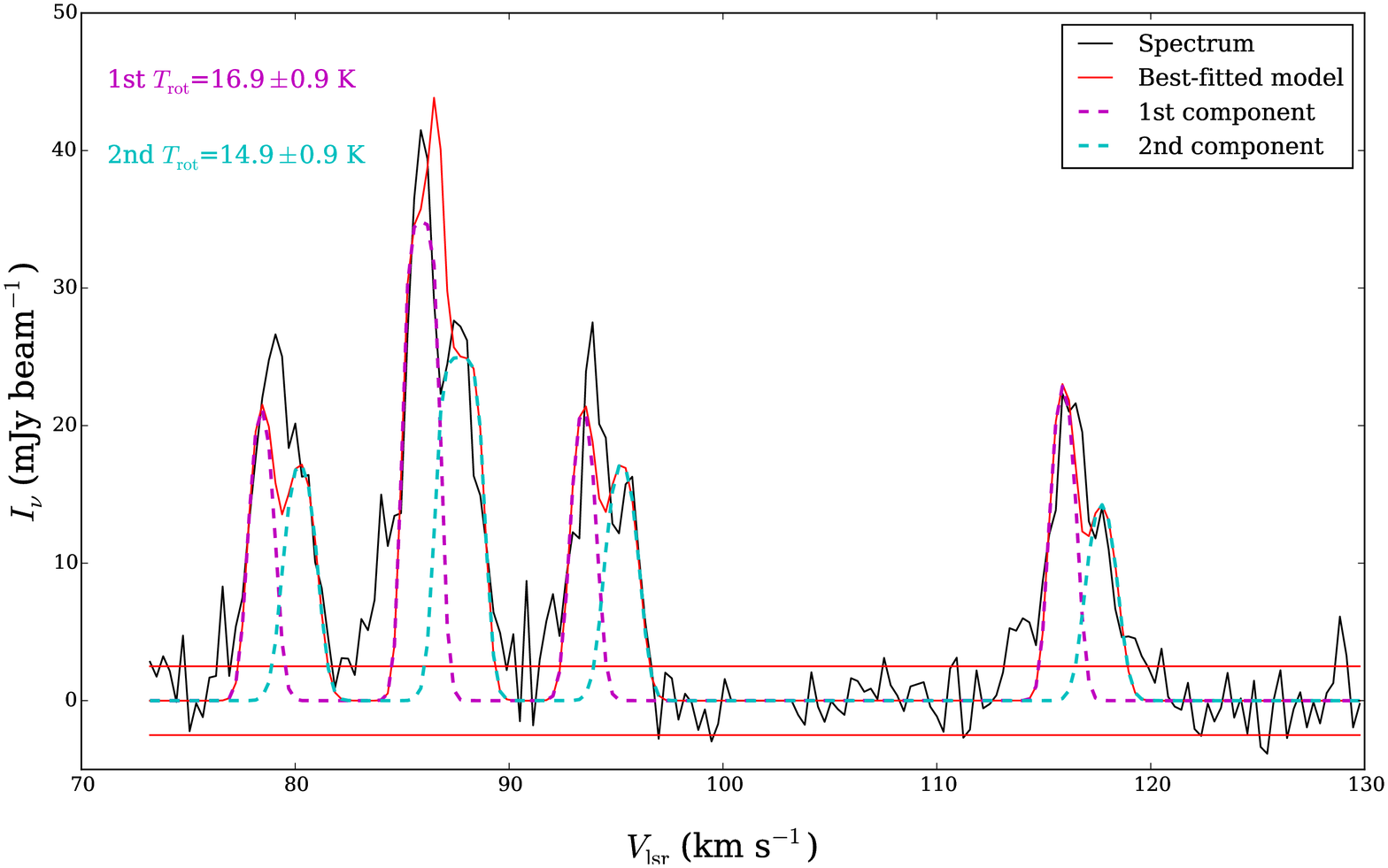} & \includegraphics[width=0.5\textwidth]{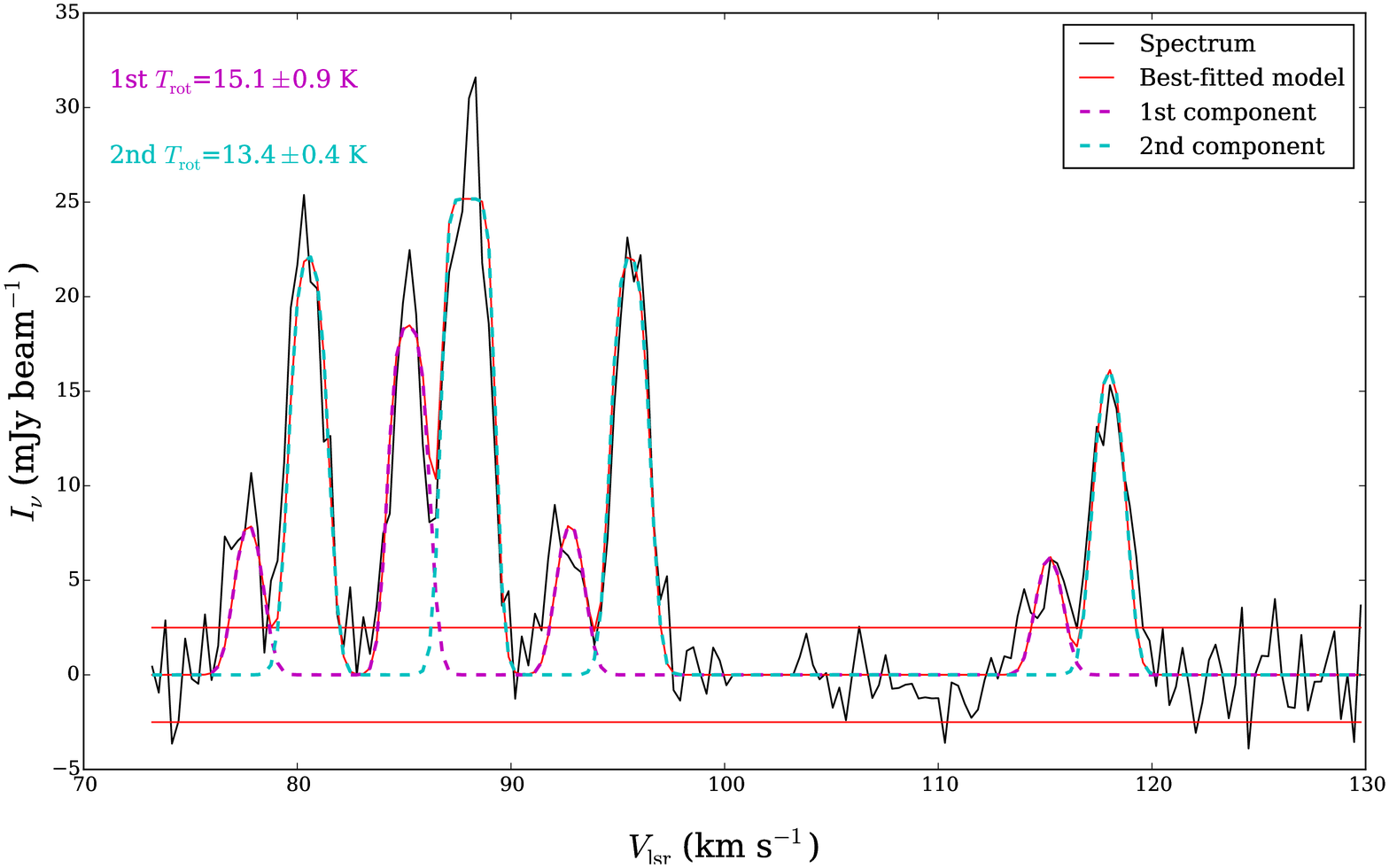} \\
\includegraphics[width=0.5\textwidth]{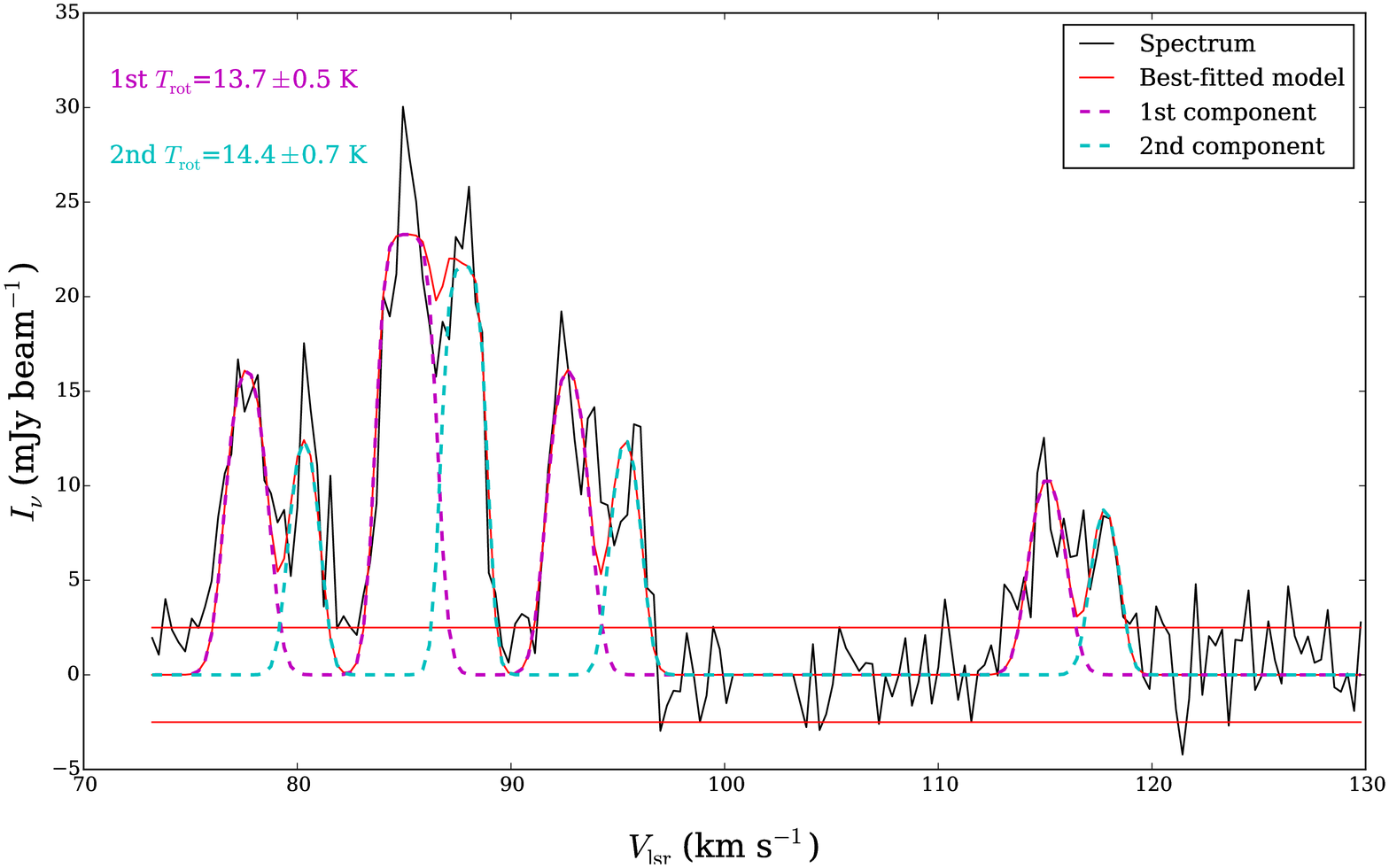} & \includegraphics[width=0.5\textwidth]{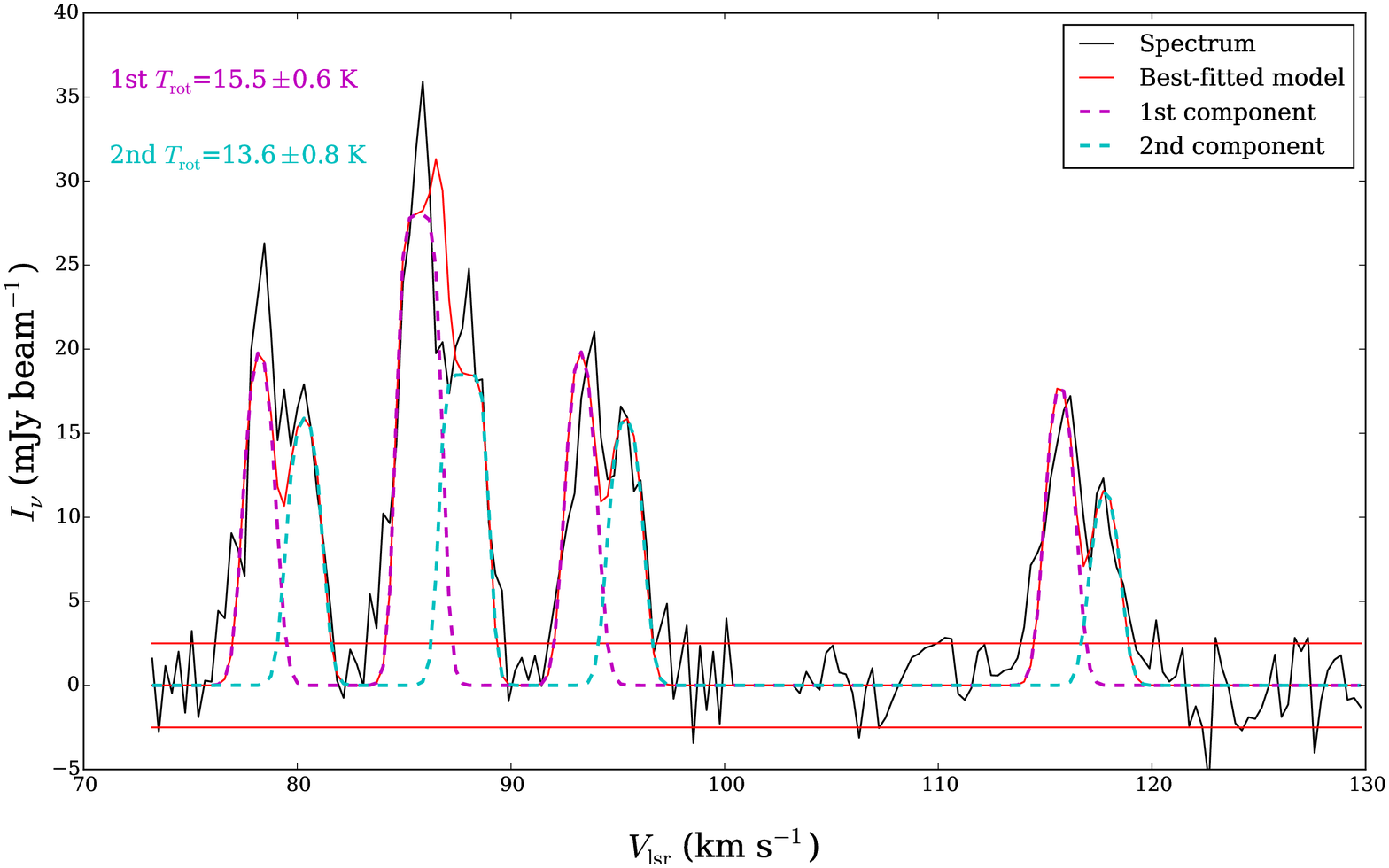} \\
\includegraphics[width=0.5\textwidth]{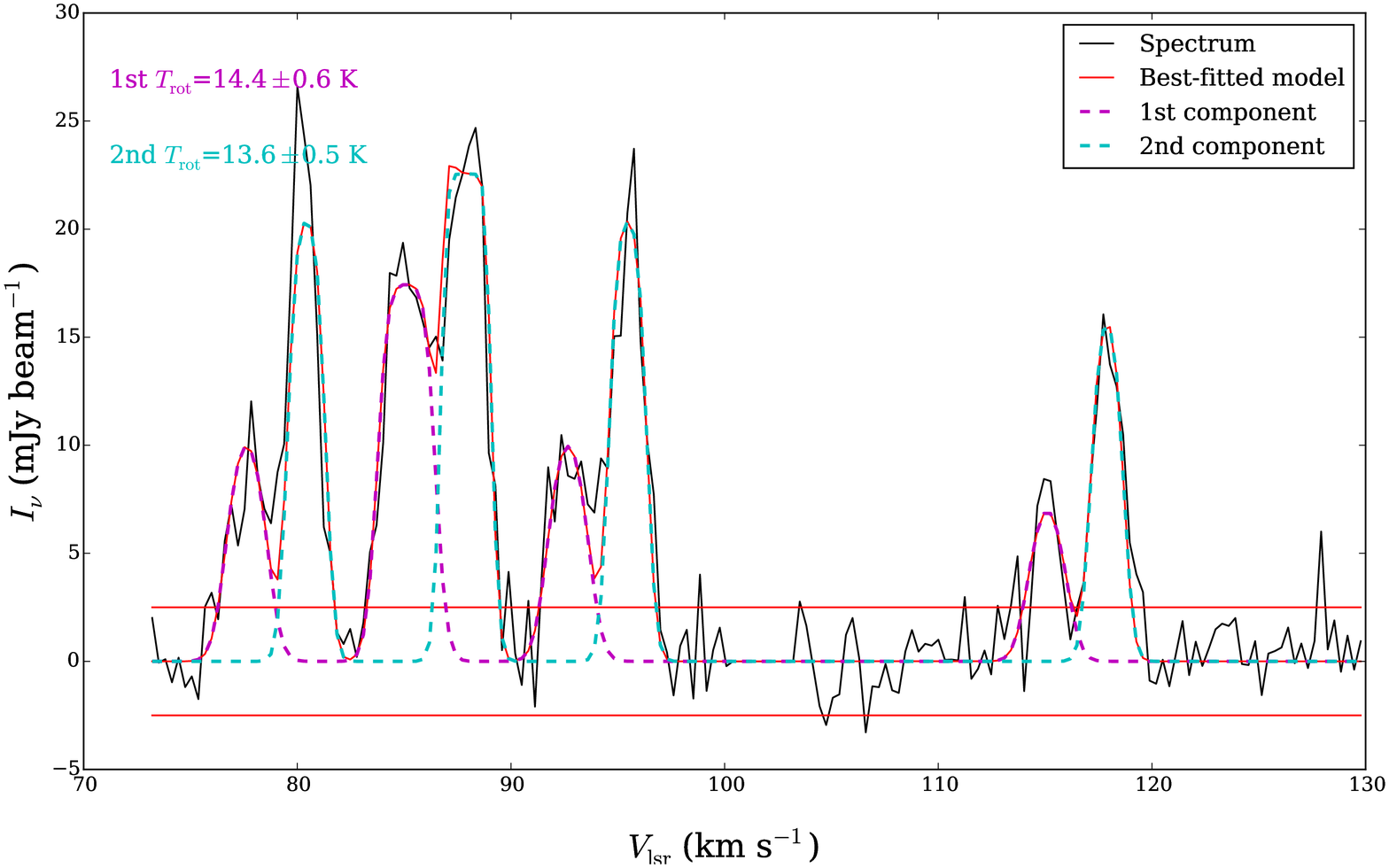} & \includegraphics[width=0.5\textwidth]{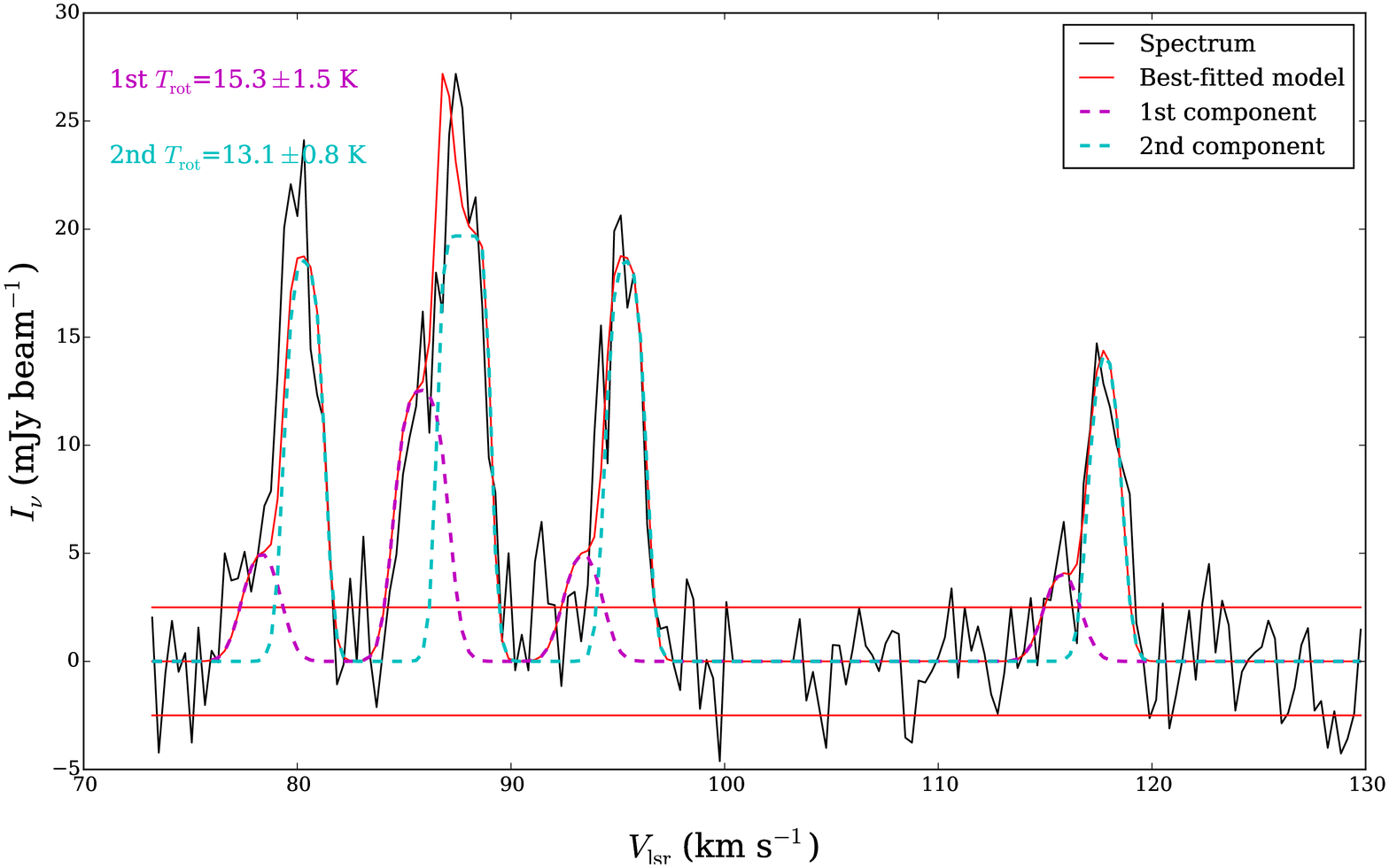}
\end{tabular}
\caption{Two-components fitting of mean spectra of the cores in MM1. The results are listed in \autoref{tab:2comp}. The \ammone{} spectra are between 70 and 100~\kms{}, while the \ammtwo{} spectra are shifted by 30~\kms{} to be between $\sim$105 and 130~\kms{}, so that they can be fitted simultaneously. For each spectrum, the dashed magenta and cyan curves represent the two components that are fitted, while the solid red curve represent the sum of them. The horizontal solid lines mark the 3$\sigma$ levels of each spectrum.}
\label{fig:2comp}
\end{figure}

\begin{figure}
\includegraphics[width=0.5\textwidth]{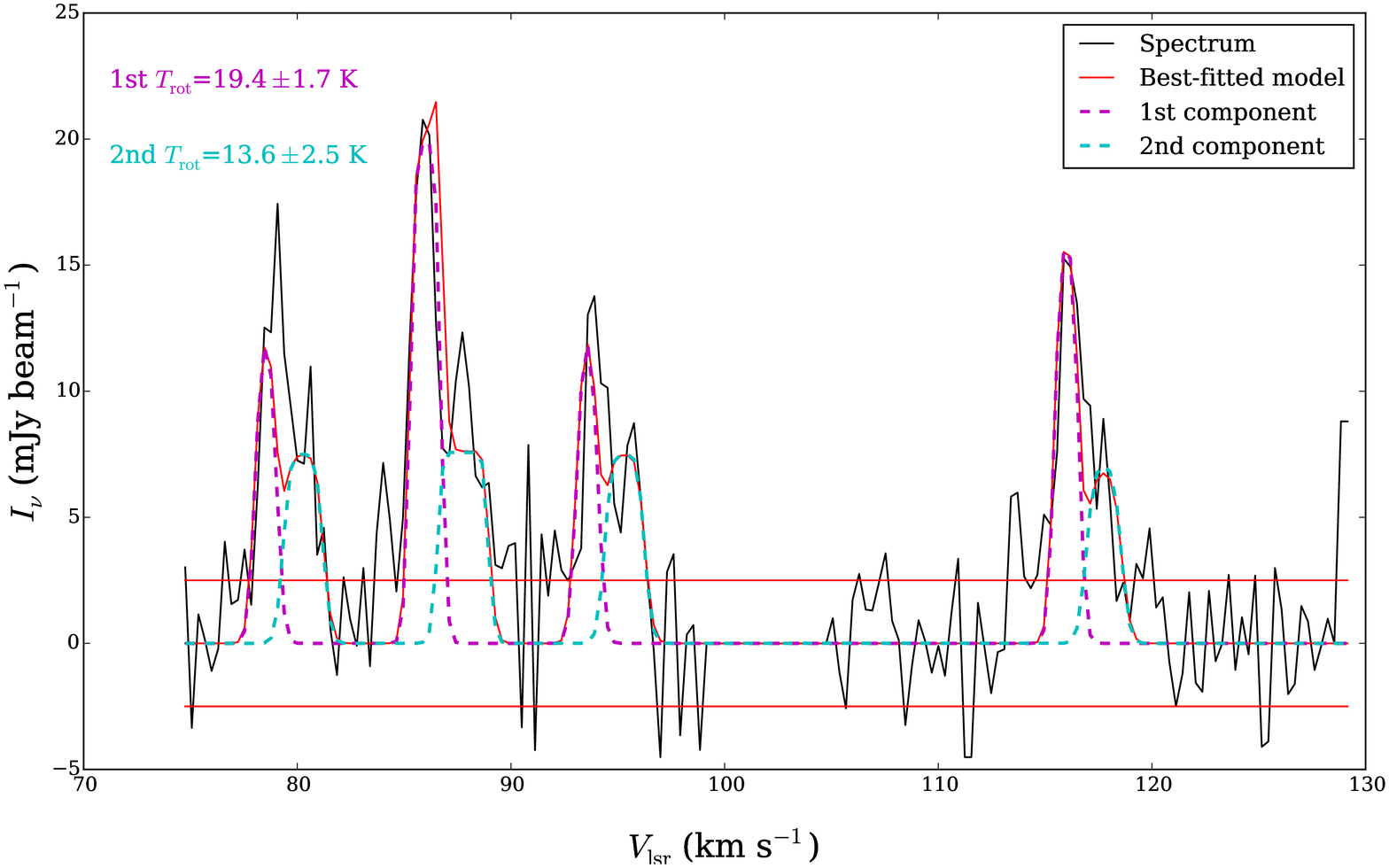}
\caption{Two-components fitting of mean spectra of MM1-p1, using the \mbox{VLA} data only. The legends are the same as in \autoref{fig:2comp}.}
\label{fig:vlaonlytemperature}
\end{figure}

\begin{figure}
\begin{tabular}{p{5.6cm}p{5.6cm}p{5.6cm}}
\includegraphics[width=0.33\textwidth]{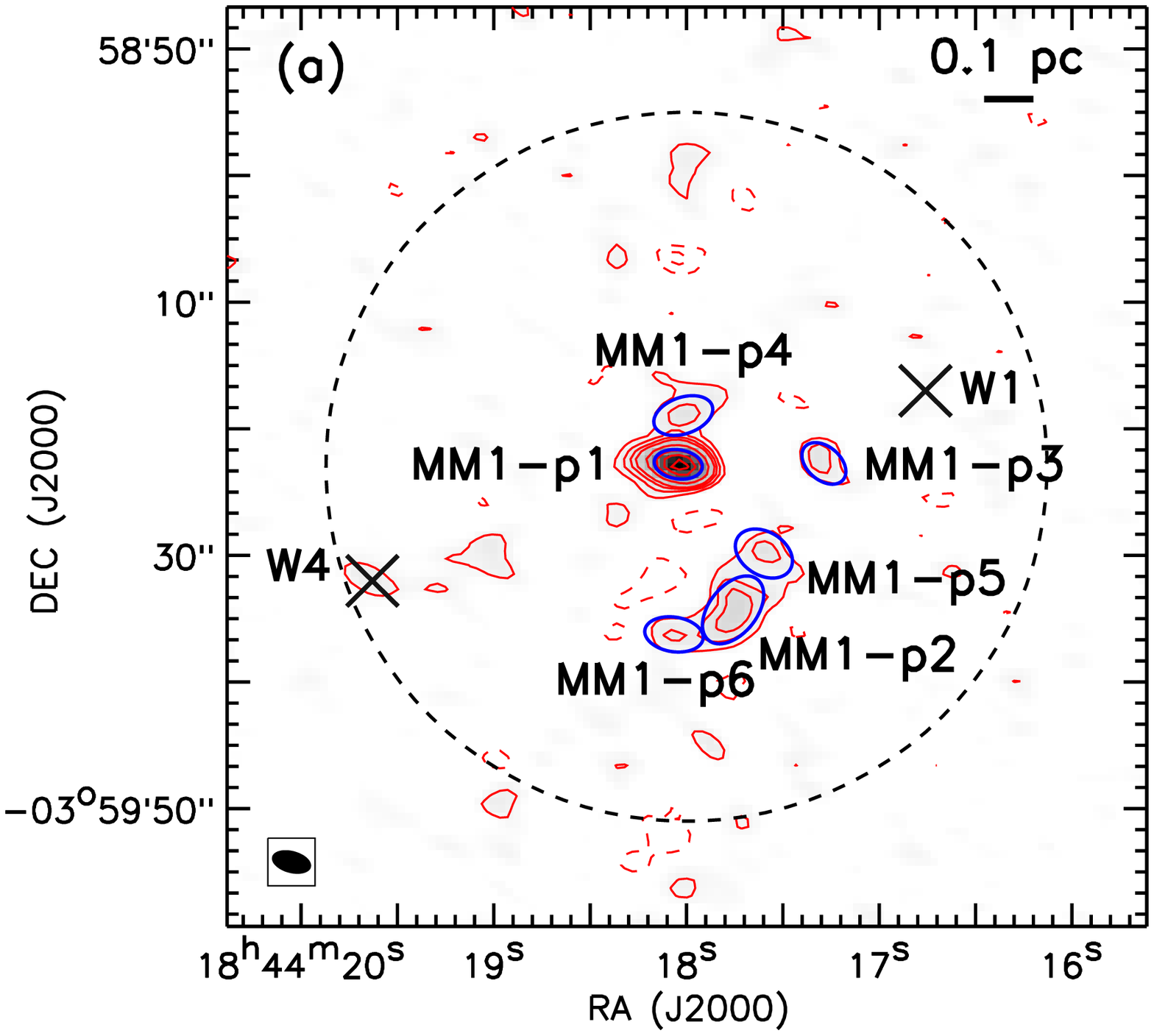} & \includegraphics[width=0.33\textwidth]{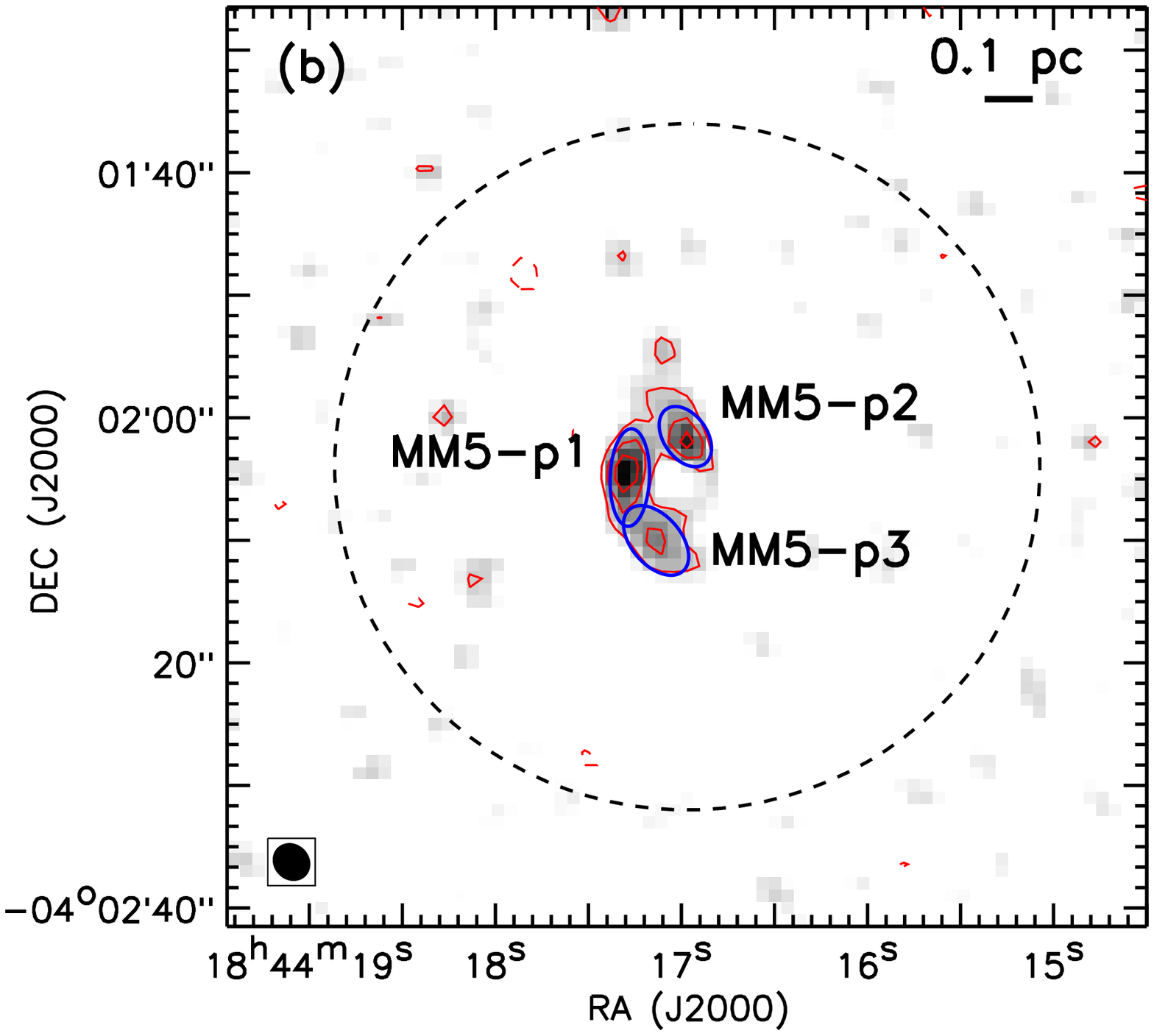} & \includegraphics[width=0.33\textwidth]{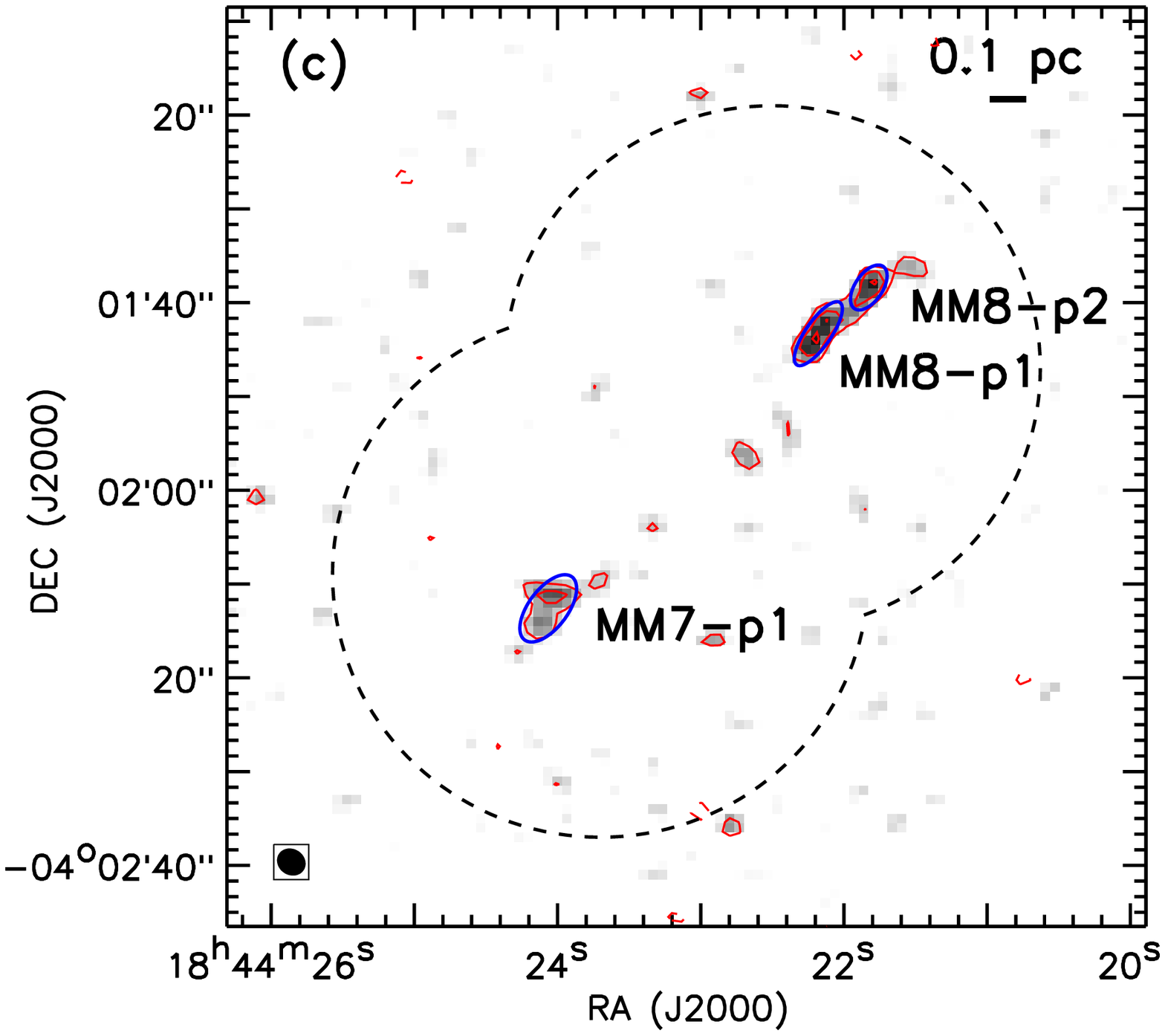}
\end{tabular}
\caption{The SMA 1.3 mm continuum emission of MM1, MM5 and MM7/8. The dashed circles in each panel mark the 56\arcsec{} primary beams of the SMA. The ellipses mark FWHM of the 2D gaussians fitted to each core. (a) The contours are in steps of 1.1~\mbox{mJy\,beam$^{-1}$}$\times$[$-$5, $-3$, 3, 5, 7, 9, 15, 25, 35, 45]. The two \water{} masers, W1 and W4, are marked with crosses. (b) The contours are in steps of 0.9~\mbox{mJy\,beam$^{-1}$}$\times$[$-3$, 3, 5, 7]. (c) The contours are in steps of 0.81~\mbox{mJy\,beam$^{-1}$}$\times$[$-3$, 3, 5, 7].}
\label{fig:sma_MMs}
\end{figure}

\begin{figure}
\centering
\includegraphics[width=0.8\textwidth]{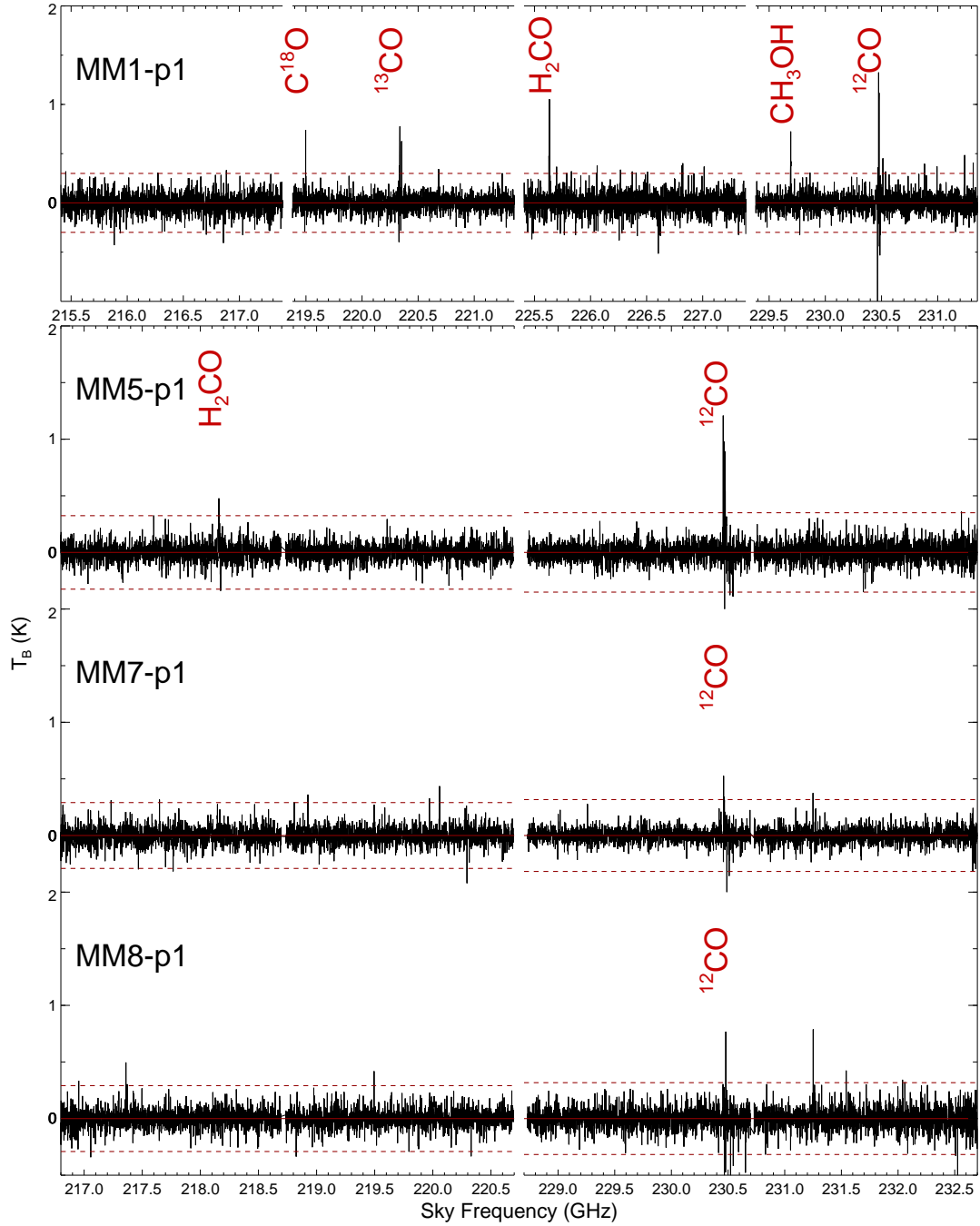}
\caption{Spectral lines of MM1-p1, MM5-p1, MM7-p1, and MM8-p1 detected by the SMA. The detected species are labelled in each panel. The horizontal dashed lines mark the 3$\sigma$ levels of each spectrum.}
\label{fig:sma_spec}
\end{figure}

\begin{figure}
\centering
\includegraphics[angle=-90,width=0.78\textwidth]{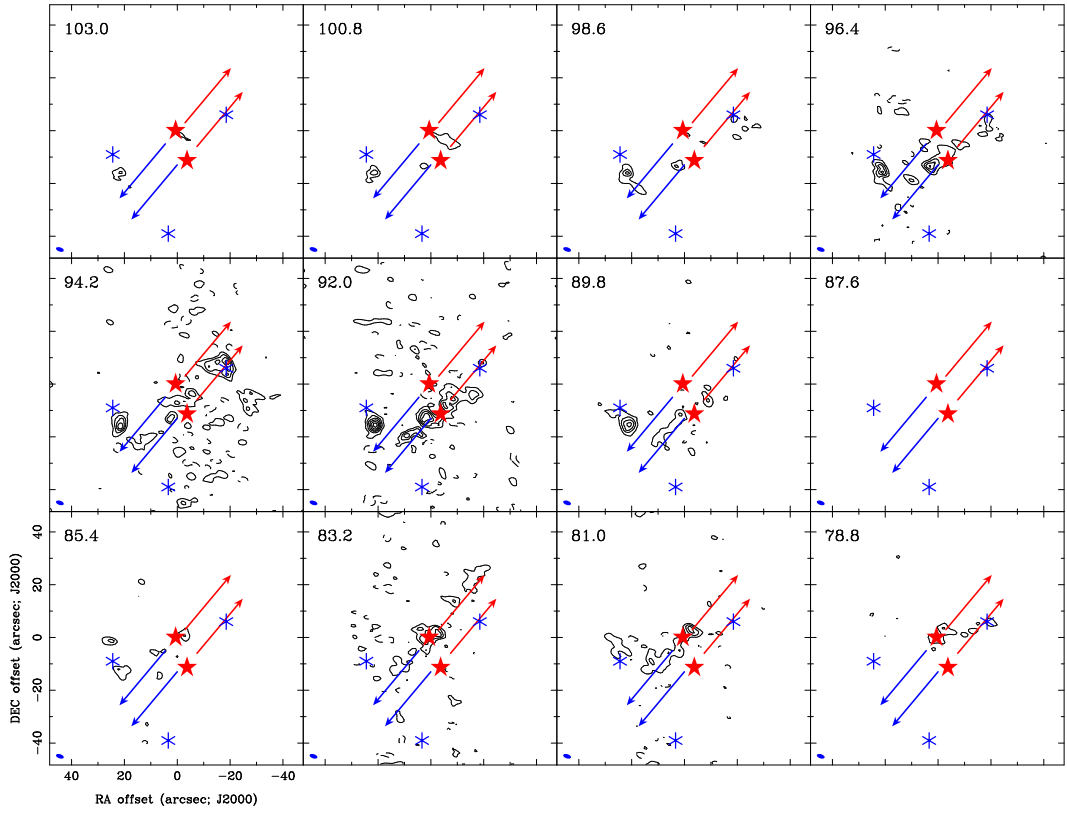}
\caption{SMA CO channel maps of MM1. The contours are in steps of 0.1~Jy\,beam$^{-1}$$\times$[-5,5,10,15,20,25,30]. MM1-p1 and MM1-p2 are marked with two stars. The two CO outflows are represented by arrows, with colors indicating blue or red shifted components. The three \water{} masers, W1, W3, and W4 are marked with asterisks.}
\label{fig:CO_MM1}
\end{figure}

\begin{figure}
\centering
\includegraphics[angle=-90,width=0.78\textwidth]{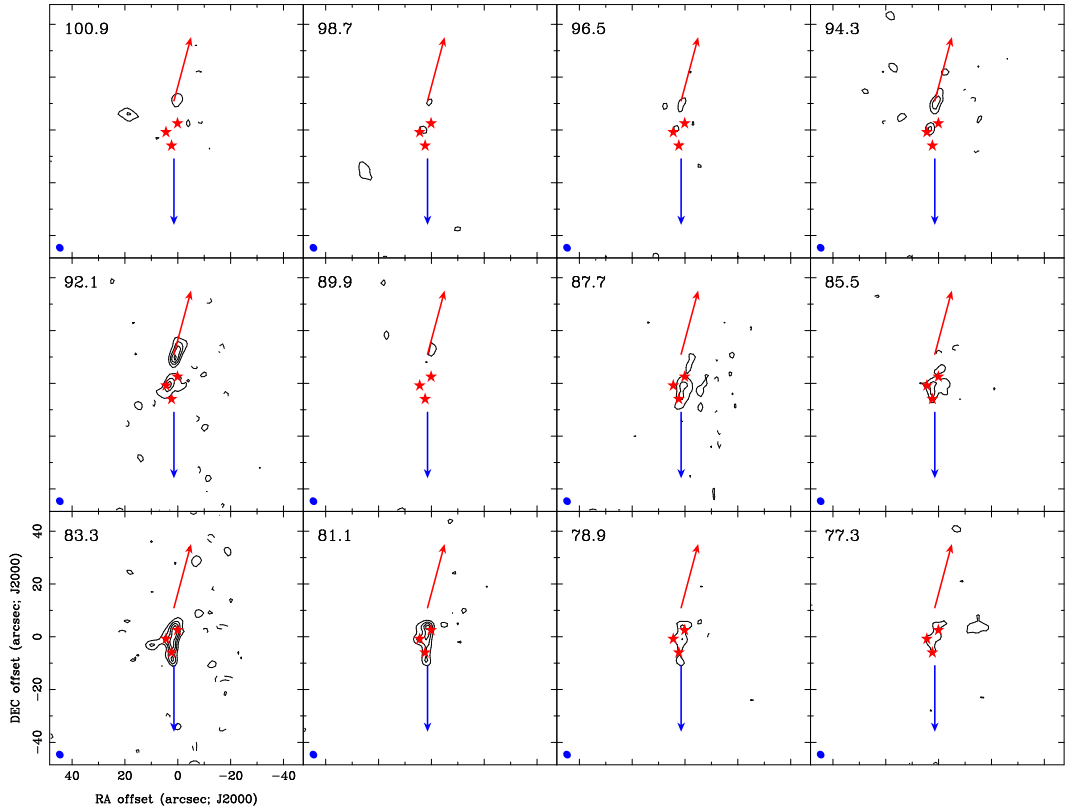}
\caption{SMA CO channel maps of MM5. The contours are in steps of 0.05~Jy\,beam$^{-1}$$\times$[-5,5,10,15]. The three cores in MM5 are marked with stars. The arrows represent the CO outflow with colors indicating blue or red shifted component.}
\label{fig:CO_MM5}
\end{figure}

\water{} masers in star forming regions are believed to be excited in shocked ambient gas \citep{elitzur1989}, therefore could trace protostellar outflows. Surveys have found \water{} masers to be associated with protostars with a wide range of luminosities \citep[e.g.][]{furuya2001,szymczak2005}. \water{} masers have been used as star formation indicators in IRDCs \citep{wang2006}. As discussed above, W1 is not likely associated with the outflow driven by MM1-p2, but is coincident with a dense core revealed by \fmh{}. It is also associated with an infrared source as shown in \autoref{fig:iram}. W2 is to the north of MM1, far away from the field of the SMA observations, so we could not relate it to any dense cores or outflows. It is spatially coincident with MM9 defined by \citet{rathborne2006}. We also found \amm{} emission as well as a 24~\micron{} infrared source associated with it in Figures~\ref{fig:iram} and \ref{fig:temperature}. W3 is also beyond the scope of the SMA observations, spatially coincident with MM4 defined by \citet{rathborne2006}. A dust core found by \citet{swift2009} is likely coincident with it, although the association is uncertain because of a lack of measurement of the core velocity. W4 is within the SMA field, and is coincident with weak dust emission as shown in \autoref{fig:sma_MMs}a. In \autoref{fig:CO_MM1}, there is a strong CO emission component to the south of W4 at 90--100~\kms{}, while only weak CO emission shows up symmetrically to the north of it at 81~\kms{}, resulting in a central velocity of $\sim$85--86~\kms{} if there was an outflow, which is consistent with the velocity of W4. Therefore, W1 and W4 might trace two protostars within MM1 that are missed by the SMA continuum observations.

\section{DISCUSSIONS}\label{sec:discussions}

\subsection{Gas fragmentation}\label{subsec:disc_fragmentation}

The SMA observations have resolved the four clumps into 12 cores in total in dust emission. We compare our result with what Jeans fragmentation predicts, in which an initially homogeneous piece of gas will fragment into smaller pieces with typical size of the Jeans length and typical mass of the Jeans mass, given a characteristic velocity and a particle number density. In our case, the four clumps fragment into cores, thus we can calculate the Jeans masses of the clumps and compare them with the core masses.

The Jeans length is defined as
\begin{equation}\label{eq:jlength}
\begin{aligned}
\lambda_\mathrm{J}&=v\left(\frac{\pi}{G\rho}\right)^{1/2}\\&=0.11\ \mathrm{pc}\left(\frac{v}{0.1\ \mathrm{\kms{}}}\right)\left(\frac{n}{10^4\ \mathrm{\cc{}}}\right)^{-1/2},
\end{aligned}
\end{equation}
where $v$ is either the sound speed $c_s$ or the velocity dispersion $\sigma_v$, and $n$ is the particle density of the clump. We used the mean \amm{} temperature of the clump to calculate $c_s$. With the parameters reported in \citet{rathborne2006,rathborne2010}, we derived the density of each clump as $n=\mathrm{mass}/(4/3\cdot\pi\cdot\mathrm{radius}^3)\times1/(\mu m_{\rm H})$, where the mean molecular mass number is $\mu=2.33$. Note that for MM5, we used the 1.2~mm flux in \citet{rathborne2006}, the dust emissivity index in \citet{rathborne2010}, and the mean \amm{} rotation temperature to calculate the clump mass using \autoref{eq:mcore}, instead of directly quoting the mass in \citet{rathborne2010}, because of the possibly incomplete spectral energy distribution (SED) fitting (see Section \ref{subsubsec:disc_mm5}). The resulting clump mass is 4 times larger. For consistency, we calculated masses of the other three clumps in the same way, and their masses are within a factor of 50\% with respect to the results of \citet{rathborne2010}. 

The Jeans mass is then the mass enclosed by a sphere with a radius of $\lambda_\mathrm{J}/2$:
\begin{equation}\label{eq:jmass}
\begin{aligned}
M_\mathrm{J}&=\frac{4\pi\rho}{3}\left(\frac{\lambda_\mathrm{J}}{2}\right)^3\\&=0.43\ \mathrm{M_\odot}\left(\frac{v}{0.1\ \mathrm{\kms{}}}\right)^3\left(\frac{n}{10^4\ \mathrm{\cc{}}}\right)^{-1/2}.
\end{aligned}
\end{equation}
The Jeans masses of the clumps are listed in \autoref{tab:cores}.

If the sound speed $c_s$ is used to calculate the Jeans parameters, which indicates that thermal pressure dominates during fragmentation, the Jeans masses are much smaller than the core masses. When the observed velocity dispersion $\sigma_v$ is applied in \autoref{eq:jmass}, the Jeans masses are then consistent with the observed core masses. Therefore, turbulent pressure is dominant over thermal pressure during 1$\rightarrow$0.1~pc scales fragmentation, which is consistent with fragmentation studies of other IRDCs \citep{zhang2009,zhang2011,wang2011,wang2014}.

\subsection{Gravitational collapse of cores}\label{subsec:disc_virial}
With the velocity dispersion $\sigma_v$ and the deconvolved radius $R$ based on \autoref{tab:cores}, we analyze virial status of the cores. The gas is in a virial equilibrium if the kinetic energy equals half of the gravitational energy. If the core mass is larger than its virial mass, it is gravitationally bound and will collapse. Note that in this framework, rotation of cores, magnetic field, or external pressure are all ignored, therefore the virial analysis is only robust in the simplest scenario.

The virial mass depends on density profile of the core \citep[e.g.][]{maclaren1988}. Assuming the radial density profile is $\rho(r)\sim r^{-2}$ as measured in IRDC cores \citep{wang2011}, then the virial mass is
\begin{equation}
M_\mathrm{virial}=\frac{3\sigma_v^2R}{G}=69.8\ \mathrm{M_\odot}\left(\frac{\sigma_v}{1\ \mathrm{\kms{}}}\right)^2\left(\frac{R}{0.1\ \mathrm{pc}}\right).
\end{equation}
If instead the density does not depend on the radius, $\rho(r)=\mathrm{const}$, then $M_\mathrm{virial}$ will be 1.7 times larger. The viral parameter is defined as $\alpha=M_\mathrm{virial}/M_\mathrm{core}$. The result is listed in \autoref{tab:cores}.

All of the 12 cores have a virial mass smaller than the core mass ($\alpha<1$). The virial parameters are between 0.09 and 0.7 in general. MM1-p1 has the smallest $\alpha$ of 0.09. Even assuming a uniform density profile, all but MM1-p3, MM7-p1, and MM8-p2 have $\alpha<1$. 

Therefore, in the simplest scenario where only gravitational and kinetic energies are considered, all cores are gravitationally bound and tend to collapse if $\rho(r)\sim\ r^{-2}$. The cores with significant star formation, MM1-p1/p2 and MM5-p1 (see Section~\ref{subsec:disc_mm15}), are strongly self-gravitating ($\alpha\lesssim0.2$). Even in the relative quiescent clumps such as MM7 and MM8, the cores are still gravitationally bound. 

\begin{figure}
\begin{tabular}{p{8cm}p{8cm}}
\includegraphics[width=0.45\textwidth]{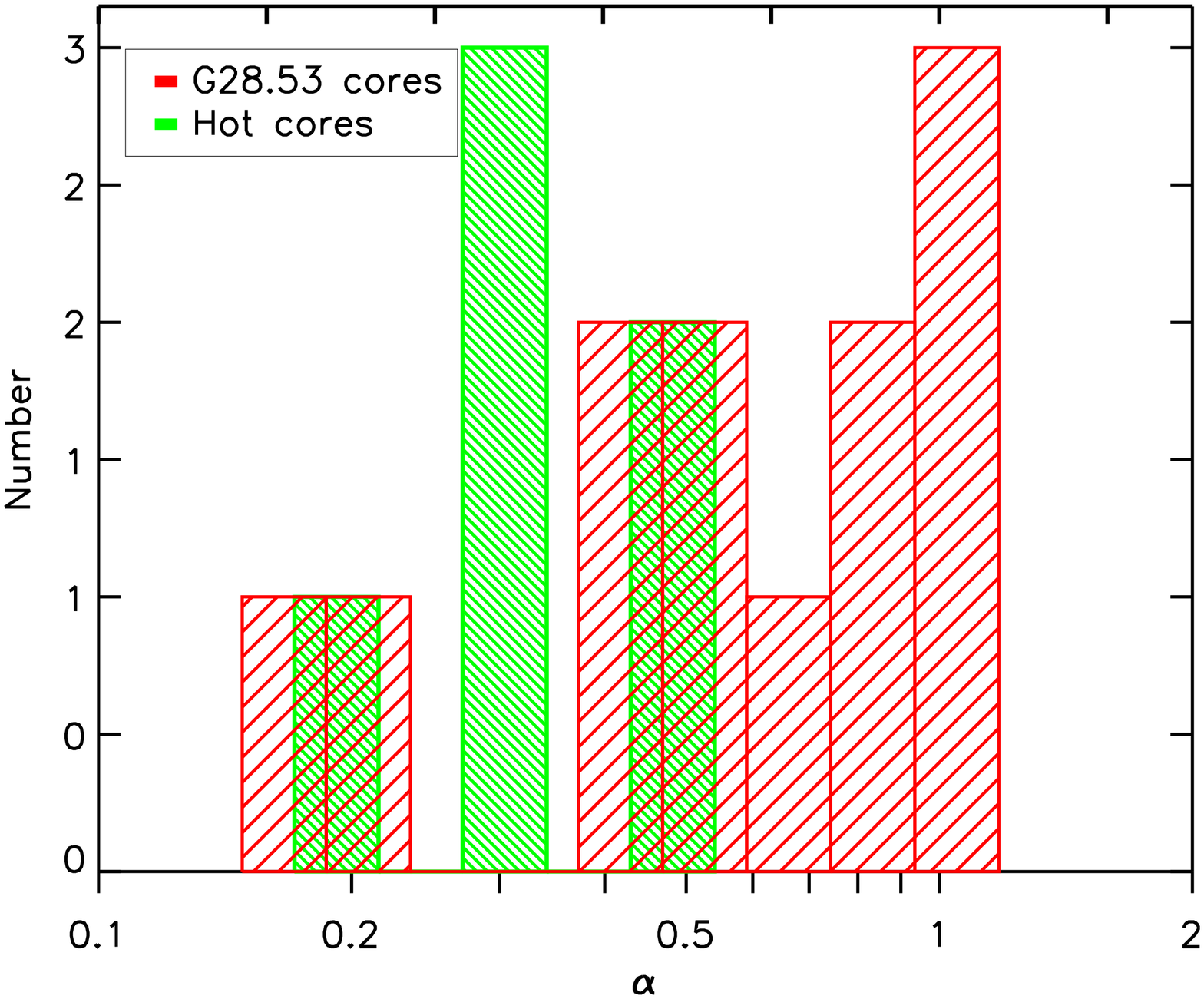} & \includegraphics[width=0.45\textwidth]{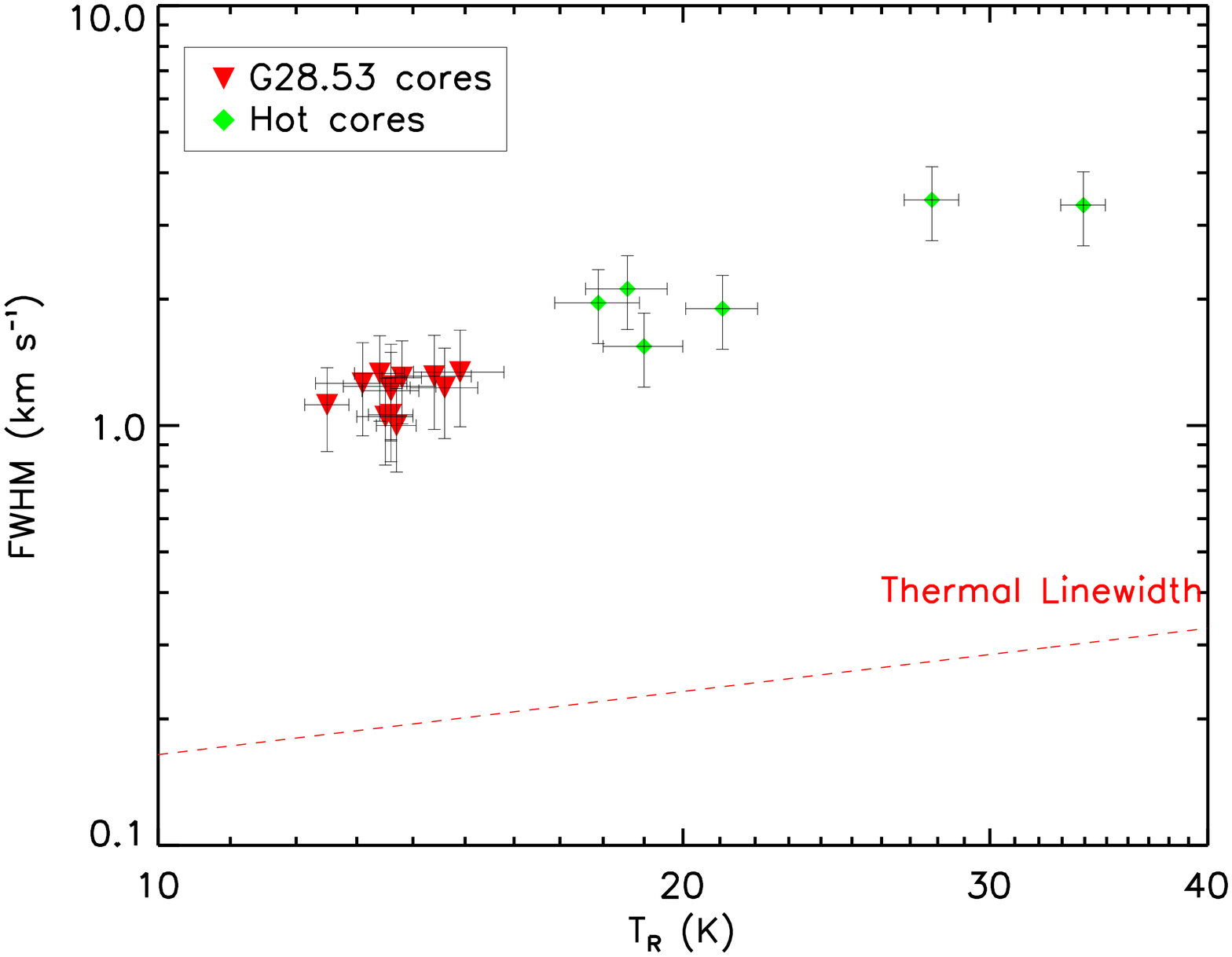}
\end{tabular}
\caption{\textit{Left}: Virial parameters of cores in G28.53 as well as in the hot core sample of \citet{lu2014}. Constant radial density profiles are assumed for all cores. \textit{Right}: FWHM linewidths vs. rotation temperatures of cores in G28.53 as well as in the hot core sample of \citet{lu2014}.}
\label{fig:nh3sample}
\end{figure}

We compared the virial parameters of the cores in G28.53 with those of hot cores in the \amm{} sample in our recent work by \citet{lu2014} in \autoref{fig:nh3sample}. The three hot core sources are IRAS~18089$-$1732, IRAS~18360$-$0537, and IRAS~18414$-$0339. Note that in \citet{lu2014}, the core masses are determined from the \amm{} column densities, although for the ones that dust continuum data are available the masses based on the two datasets are fairly consistent. The rotation temperatures of the hot cores measured from the first two inversion transitions of \amm{} should be treated as the lower limits, while the real temperatures at central part of the hot cores could be $>$300~K.

Assuming a uniform radial density profile, the virial parameters of the hot cores are all smaller than 1, similar to what we find in G28.53. The gas temperatures of the hot cores are usually $>$30~K, thus their thermal linewidths are larger than those of G28.53. However as shown in \autoref{fig:nh3sample} the non-thermal linewidths are always dominant over the thermal linewidths, thus the increase of linewidths by a factor of 2--3 from G28.53 cores to hot cores is mainly due to the non-thermal motions. Therefore, the fact that the virial parameters are consistent from G28.53 cores to hot cores indicates that turbulence is essential in supporting the cores against gravity constantly in the evolution of dense cores. Given that the hot cores are more massive than the IRDC cores (typically a few hundreds of M$_\odot$), to maintain a similar virial parameter, turbulence must be enhanced. The feedback from protostars in hot cores, including outflows and heating, could increase the gas temperatures, and more importantly, inject turbulent energy into gas, which maintains a steady accretion of dense gas to form high-mass stars.

As one of the most important energy sources against gravity other than kinetic energy, magnetic field is not considered since we do not have measurements in this cloud. If the typical magnetic field strength $B=1$~mG in high-mass star formation regions \citep{cortes2006,girart2013,qiu2013,qiu2014,frau2014,zhang2014} is added into the virial relation, and assuming a uniform radial density profile, then
\begin{equation}
(3\sigma_v^2+\frac{1}{2}\sigma_\textrm{A}^2)M-\frac{3}{5}\frac{GM^2}{R}=0,
\end{equation}
in which $\sigma_\textrm{A}=B/\sqrt{4\pi\rho}$ is the Alfv\'{e}nic velocity. The total virial mass $M$ is 
\begin{equation}
M=M_\mathrm{virial}+M_\mathrm{B}=\frac{5(\sigma_v^2+\sigma_\textrm{A}^2/6)R}{G},
\end{equation}
where $M_\mathrm{virial}=5\sigma_v^2R/G$ and 
\begin{equation}
M_\mathrm{B}=5\sigma_\textrm{A}^2R/6G=\left(\frac{M_\mathrm{mag}}{M_\mathrm{core}}\right)M_\mathrm{mag}
\end{equation}
is the magnetic virial mass. Here
\begin{equation}
M_\mathrm{mag}=\frac{R^2B}{\sqrt{18G/5}}
\end{equation}
is the traditionally defined critical mass for a spherical cloud of uniform density with a uniform magnetic field \citep{strittmatter1966}. Only when $M_\mathrm{mag}=M_\mathrm{core}$ does the magnetic virial mass $M_\mathrm{B}$ we defined equal $M_\mathrm{mag}$.

In this case, the virial parameters of MM1-p1/p2 and MM5-p1 are still smaller than 1, while all the other cores have $\alpha>1$. This might explain why signatures of star formation are only detected in these three cores (see Section \ref{subsec:disc_mm15}). Therefore, magnetic field might also be an important factor in determining the dynamics of the cores.

\subsection{Massive star formation in MM1 and MM5}\label{subsec:disc_mm15}

\subsubsection{An outflow-disk system in MM1-p1}\label{subsubsec:disc_mm1p1}
In several previous studies \citep{rathborne2008,rathborne2010,swift2009,sanhueza2012}, MM1 was classified as `quiescent' with no signs of high-mass star formation, given the lack of molecular lines and absence of associated infrared sources. Our observation suggests that MM1 is not quiescent, but is already forming stars. In MM1, the two most massive cores are associated with outflows. The two H$_2$O masers W1 and W4 could also trace star formation. In total there are at least four star forming sites.

With a mass of 56.0~\msol{}, MM1-p1 is the most massive core we resolved in G28.53. It remains unresolved in the 345~GHz SMA observations at an angular resolution of $2\arcsec\times1.1\arcsec$ \citep{swift2009}, and only one \twelveco{} outflow was detected, both of which suggest there might be a single protostellar object embedded in MM1-p1, which will form a star or a multiple stellar system. The large core mass and the energetic outflow suggest high-mass star formation. The small virial parameter of 0.09--0.2 suggests that it is strongly self-gravitating.

\begin{figure}
\begin{tabular}{p{8.5cm}p{8.5cm}}
\includegraphics[width=0.45\textwidth]{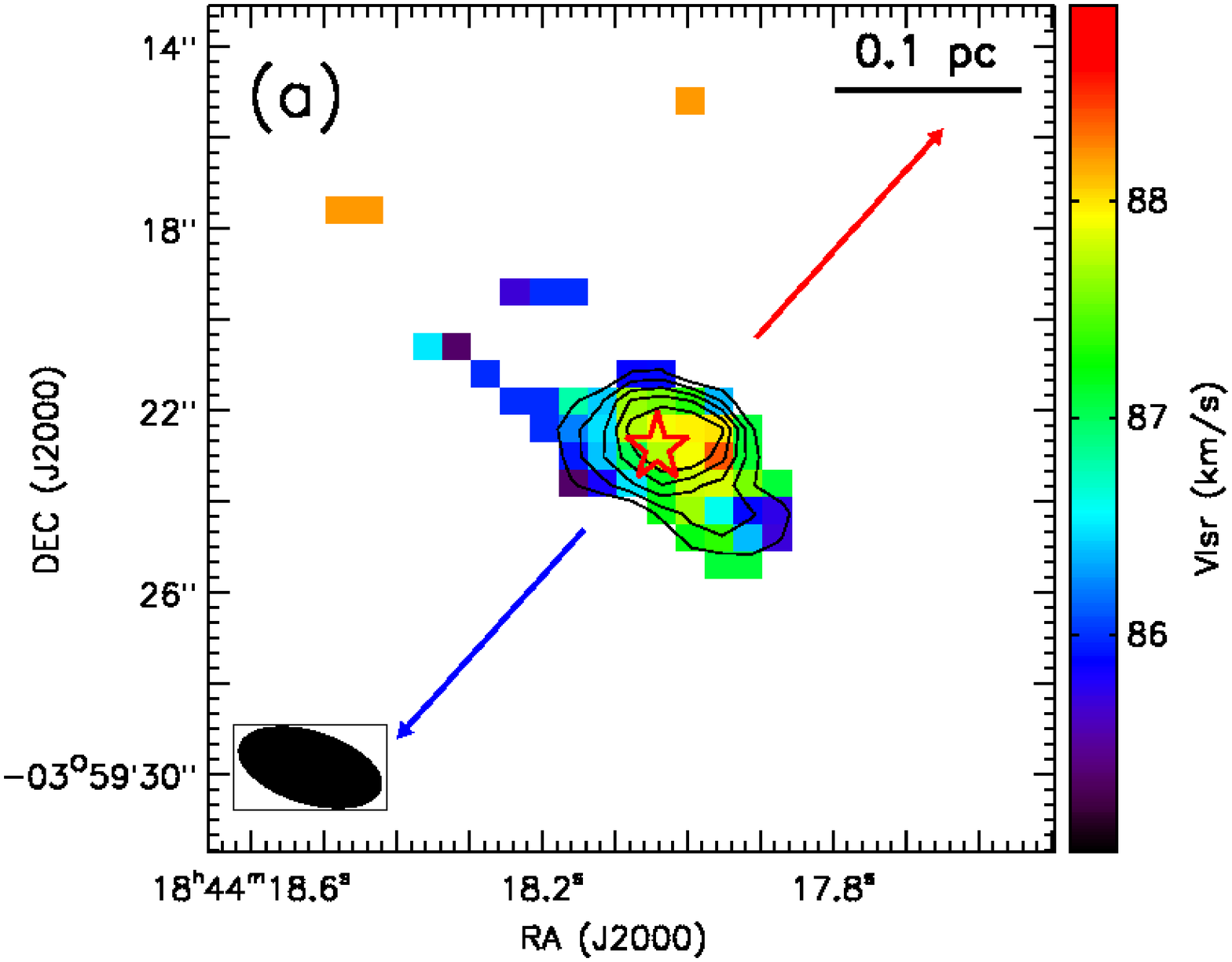} & \includegraphics[width=0.45\textwidth]{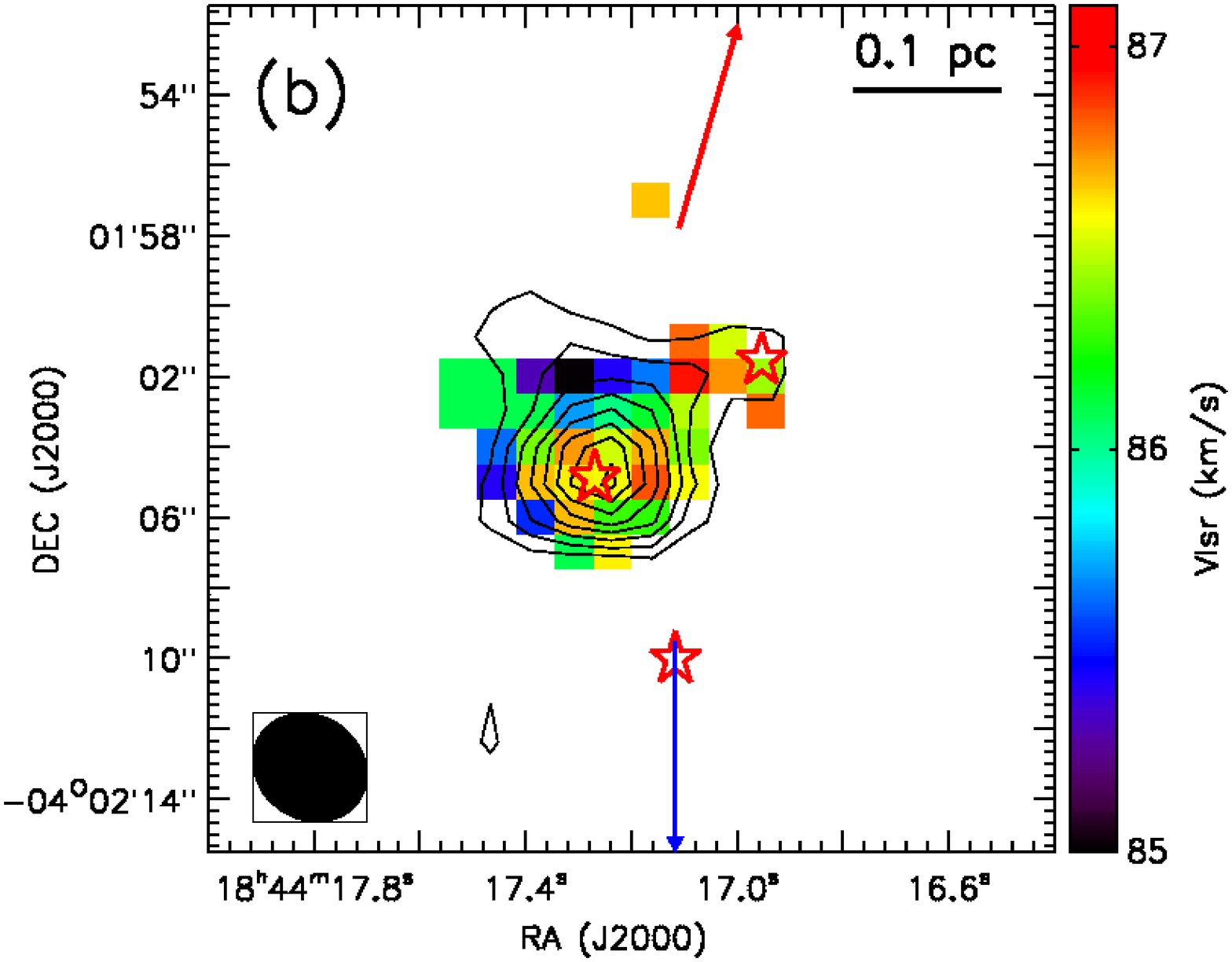}
\end{tabular}
\caption{(a) SMA \methanol{} velocity field of MM1-p1, overlaid with its integrated intensity. The contours are in steps of 0.13~\mbox{Jy\,beam$^{-1}$\,\kms{}}$\times$[3, 5, 7, 9, 11]. The star marks MM1-p1. The red and blue arrows represent the CO outflow associated with MM1-p1 as in \autoref{fig:CO_MM1}. (b) SMA \fmh{} velocity field of MM5, overlaid with its integrated intensity. The contours are in steps of 0.1~\mbox{Jy\,beam$^{-1}$\,\kms{}}$\times$[3, 5, 7, 9, 11, 13, 15, 17]. The three stars mark MM5-p1--p3. The red and blue arrows represent the CO outflow in MM5 as in \autoref{fig:CO_MM5}.}
\label{fig:smavfield}
\end{figure}

The SMA \methanol{} line emission at $\sim$87 \kms{} in MM1-p1 presents a flatten morphology perpendicular to the CO outflow, with a velocity gradient of $\sim$3~\kms{} across $\sim$0.1~pc, shown in \autoref{fig:smavfield}a. With a critical density of $\sim$10$^5$~\cc{}, the \methanol{} line emission is tracing a dense gas component, likely a rotating envelope surrounding a protostellar disk. These results suggest that there might be an outflow-disk system within MM1-p1.

\subsubsection{Star formation activity associated with MM1-p2 and cluster formation in MM1}\label{subsubsec:disc_mm1p2}
As \citet{rathborne2008} pointed out, MM1-p2 fragments into three condensations among which the most massive one might form high-mass stars, based on the broad line wings in singled-dish spectra of the outflow tracer SiO. With the SMA 230~GHz observations we detect the outflow directly, although it is uncertain which one of the three condensations in \citet{rathborne2008} is driving this outflow.

\begin{figure}
\begin{tabular}{p{8.5cm}p{8.5cm}}
\includegraphics[width=0.45\textwidth]{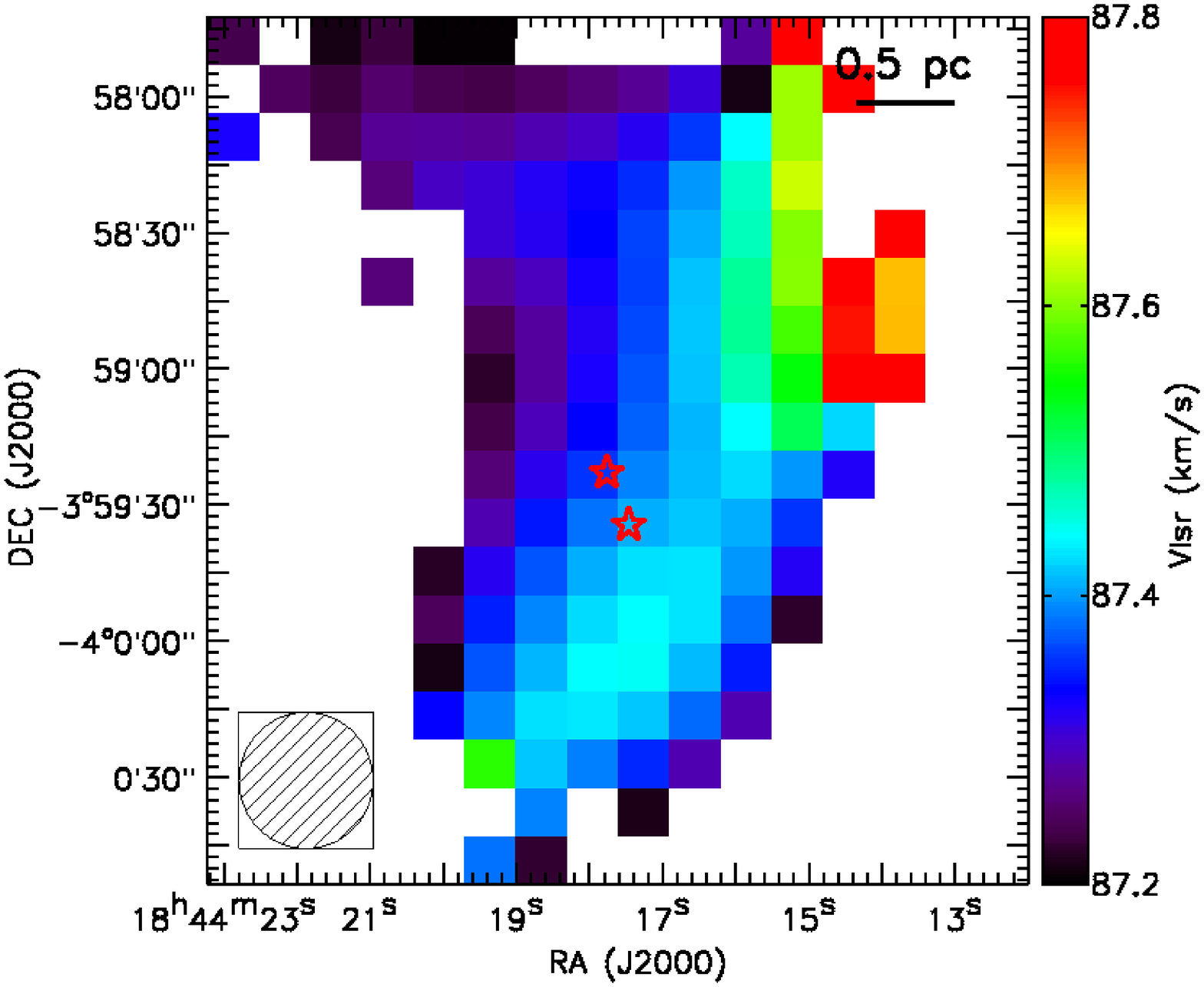} & \includegraphics[width=0.45\textwidth]{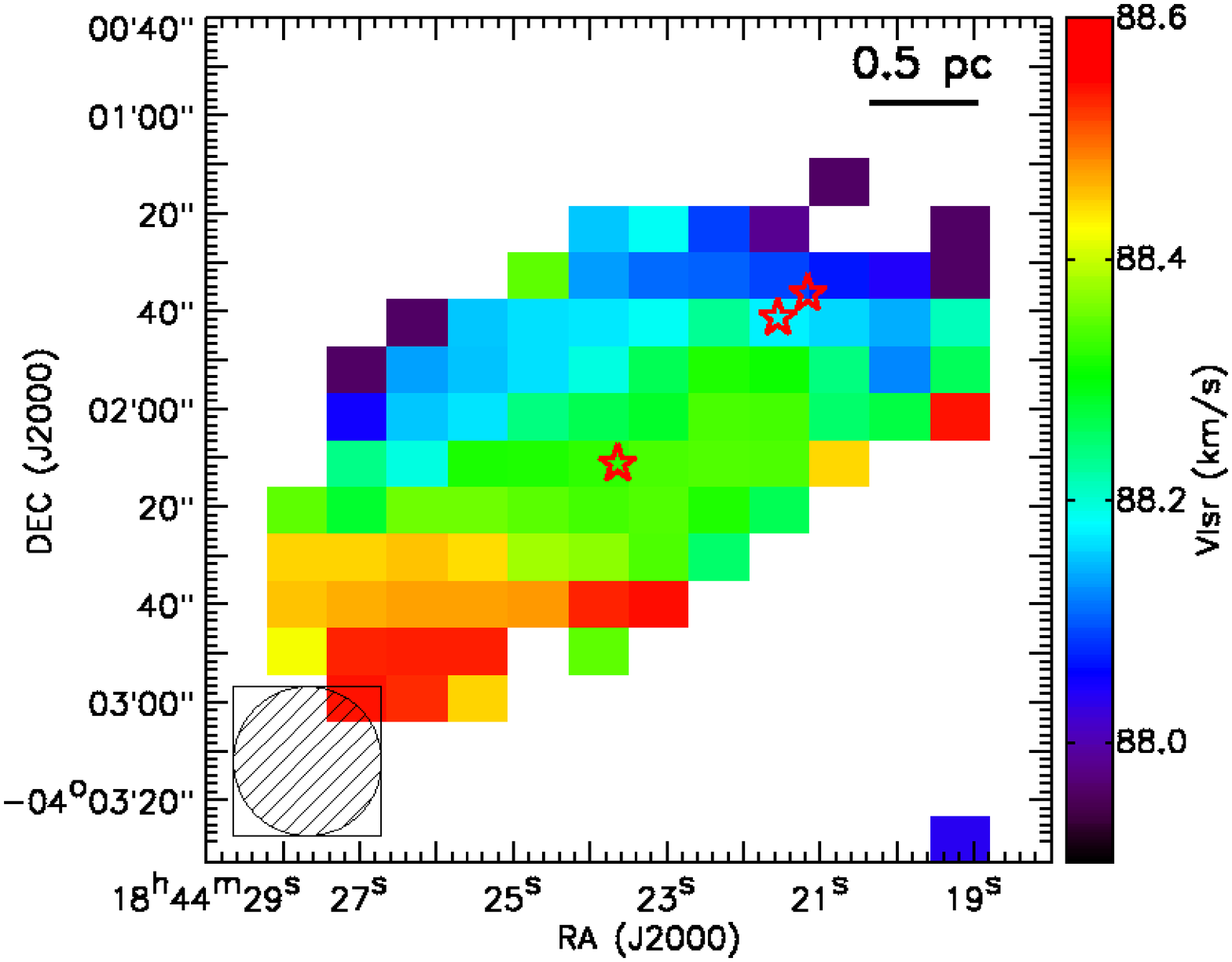} \\
\end{tabular}
\caption{GBT \ammtwo{} velocity field of MM1 (the 87~\kms{} component) and MM7/8. The 30\arcsec{} beam of GBT at K band is shown in the lower left corner in each panel. {\it Left}: MM1-p1/p2 are marked with stars. {\it Right}: MM7-p1 and MM8-p1/p2 are marked with stars.}
\label{fig:nh3vfield}
\end{figure}

The outflows associated with MM1-p1/p2 are parallel to each other. Similar nearly parallel outflows have been found in other IRDCs \citep[e.g.][]{wang2011,wang2014}. A possible interpretation is that the cores are fragmented from an initially rotating clump, then their rotation axes will tend to be parallel to the initial rotation axis of the clump given conservations of angular momentum, which leads to parallel disks and finally parallel outflows. As shown in \autoref{fig:nh3vfield}, the velocity field of the 87~\kms{} component of GBT \ammtwo{} of MM1 presents a gradient at a position angle of 40--50\arcdeg{} north to east, perpendicular to the outflows. If this velocity gradient is interpreted as rotation, then the MM1 clump is likely spinning slowly, with an axis parallel to the outflow orientations, which supports the rotating clump fragmentation scenario. However, note that such velocity gradient can also be reproduced with random motions \citep{burkert2000}. In addition, magnetic field in the clumps might also be important in aligning the outflows \citep[e.g.][]{zhang2014}, by regulating the 0.1~pc scale magnetic fields in the cores thus affecting the orientation of the accretion disks.

The two \water{} masers in MM1 could pinpoint another two star forming sites. Both are associated with weak (3$\sigma$) 345~GHz dust emission in the SMA map of \citet{swift2009}. Between them, W1 is more interesting because of its association with \fmh{} line emission and an infrared source. Three more cores were detected at levels of $>$6$\sigma$ in the southern part of MM1 in the SMA 345~GHz dust emission map of \citet{swift2009}. Therefore, with at least eleven cores, among which four are forming stars, MM1 will probably form a star cluster. MM1-p1 and the most massive condensation in MM1-p2 are two promising high-mass star formation sites.

In \autoref{fig:iram}, MM1 is the most infrared extinct part in G28.53, and it remains dark up to 70~\micron{} \citep{molinari2010}, which leads to a low luminosity \citep[e.g.\ 200~\lsol{} in][]{rathborne2010}. In contrast, MM5 starts to be bright in 24~\micron{} and has a larger luminosity \citep{rathborne2010}. The discrepancy between the more active star formation and the smaller luminosity of MM1 could be due to heavy extinction: if the 85~\kms{} component was at the front then it could extinct infrared emission produced by star formation in the 87~\kms{} component behind it. Even if the 87~\kms{} component is not shielded, the self-extinction \citep[$A_V$ up to 50--100 given a column density of $\gtrsim$10$^{23}$~\cd{},][]{guver2009} could keep it infrared dark. This is similar to the IRDC G30.88+0.03 in \citet{zhang2011}, which has a small luminosity of 460~\lsol{} but presents \water{} masers and massive cores, and there are also two velocity components along the line of sight. Therefore, infrared emission only is not sufficient in assessing star formation activity, especially in the presence of multiple velocity components, while spectral lines like \methanol{} can be used to resolve multiple components and trace deeply embedded star formation.

\subsubsection{Early phase star formation in MM5}\label{subsubsec:disc_mm5}
Previous studies \citep{rathborne2010,sanhueza2012} classified MM5 as `intermediate', based on infrared emission and chemistry, e.g.\ it contains a 24~\micron{} source but not a green fuzzy at 4.5~\micron{}. Our results support this conclusion, with MM5 in an `intermediate' evolutionary phase between the `active' MM1 and the `quiescent' MM7/8.

A bipolar outflow in the north-south direction is found in MM5, while the angular resolution is not good enough to tell which of the three cores it originates from. All three cores are self-gravitating, with masses between 24 and 38~\msol{}, thus are equally possible to harbor protostars. The outflow is less energetic and slightly younger than the two outflows in MM1, suggesting that the star formation in MM5 is less active and possibly at an earlier phase than in MM1.

With a critical density of $\gtrsim$10$^5$~\cc{} \citep{guzman2011}, \fmh{} usually traces dense gas. As shown in \autoref{fig:smavfield}b, the SMA detected \fmh{} line emission associated with MM5-p1/p2, which is elongated in the east-west direction and presents a velocity gradient across MM5-p1 and MM5-p2, perpendicular to the outflow. Therefore, the \fmh{} emission is likely tracing a dense envelope that encloses the two cores, and perhaps spins slowly, which might feed material into disks around embedded protostars. The two cores, MM5-p1/p2, are hence the potential driving sources of the outflow.

The \amm{} rotation temperatures in MM5, using either combined data or VLA data only, are generally $\lesssim$15~K. We did not find any signature of heating associated with outflows or hot cores. The protostars in the three cores, if there are any, must be at an early stage when accretion or stellar luminosity is not strong enough to heat the ambient gas. Note that \citet{rathborne2010} derived a dust temperature of 30~K and a luminosity of $\gtrsim$10$^3$~\lsol{} for MM5, from SED modeling. The robustness of this result suffers from the fact that they used fluxes in only four bands (24~\micron{}, 450~\micron, 850~\micron, 1.2~mm), making the modeling less constrained at around 100~\micron{} which is supposed to be the peak of the SED. Recently \citet{schneider2014} fitted the SED of G28.53 using Herschel 150--500~\micron{} data and derived a dust temperature of 16--18~K for MM5, consistent with our \amm{} rotation temperature, which is expected since at high densities ($\gtrsim$10$^{4.5}$~\cc{}) the gas and dust temperatures are usually well coupled \citep{goldsmith2001}. The results of MM1 and MM7 which include data from more bands in \citet{rathborne2010} are in general consistent with ours.

With a total mass of $\sim$400~\msol{} \citep{rathborne2006,rathborne2010}, MM5 has sufficient dense gas to form $\sim$80~\msol{} stars assuming a star formation efficiency of 20\%. The 24~\micron{} point source, the outflow, the spinning \fmh{} envelope, and the self-gravitating cores all suggest the capability of high-mass star formation, while the absence of hot core tracers such as \methanol{} and the low temperature indicate little stellar feedback. 

\subsection{Potential of star formation of MM7 and MM8}\label{subsec:disc_mm78}

The three cores in MM7/8 are marginally gravitationally bound, and do not show any star formation signature such as outflow, maser, infrared source or increased temperature. As previous studies suggested \citep{rathborne2010,sanhueza2012}, these two clumps are likely `quiescent', with no current star formation.

However, given that the cores are self-gravitating, MM7/8 still have the potential of collapsing and eventually forming stars. Their masses are $\sim$300 and $\sim$240~\msol{} respectively using the fluxes and $\beta$ reported in \citet{rathborne2010}, which are comparable to the mass of MM5. With mean densities of 1$\times$10$^4$--2$\times$10$^4$~\cc{}, the free fall time scale is 0.2--0.4~Myr. If the cores are indeed collapsing even after taking magnetic field into account, then star formation will start in such a time scale.

It is also worth noting that MM7/8 are embedded in a filamentary structure, which presents a velocity gradient of 1~\kms{} along the major axis, which is consistent with a gas stream falling into the main part of G28.53, evidenced by the GBT \ammtwo{} velocities in \autoref{fig:nh3vfield}. In fact, G28.53 is the densest and coldest part of a large filament which extends $\sim$60~pc (Wang et al. submitted), in which MM7/8 connect the main part of G28.53 and the large filament. Assuming a proper motion velocity of $\sim$1--2~\kms{} and a projected distance of 2--3~pc, MM7/8 will collide with MM1 in 1--3~Myr. For reference, the free fall time scale of the main part of G28.53 \citep[MM1/2/3/4/6, $\sim$3.5$\times$10$^3$~\msol{} within 3~pc;][]{rathborne2010} is $\sim$1.4~Myr. Therefore MM7/8 will probably first form stars, then dynamically interact with the other clumps at the center of G28.53 due to their gravitational pull. This picture is similar to what \citet{baobab2012a} proposed for the high-mass cluster formation in G10.6$-$0.4, in which fragmentation occurs in filaments on free fall time scales, while the filaments themselves fall into a central dense region where OB clusters are forming through a global contraction.

\section{CONCLUSIONS}\label{sec:conclusions}

The multi-frequency observations of the IRDC G28.53 have revealed twelve 0.1~pc scale cores in four massive clumps. We obtain gas temperatures and kinematics from VLA and GBT \amm{} lines, and derive core masses based on the SMA dust continuum emission. We also assess star formation associated with these cores using VLA masers and the SMA spectral lines. Our conclusions are:
\begin{enumerate}
\item When 1~pc scale clumps in IRDCs fragment into 0.1~pc scale cores, turbulent pressure is more important than thermal pressure in supporting the gas. Therefore, it is possible to form super-thermal-Jeans-mass cores that are massive enough to harbor high-mass stars.
\item All of the cores we find in IRDC G28.53 have a virial parameter $\alpha$ ($M_\mathrm{virial}/M_\mathrm{core}$) smaller than 1 assuming a radial density profile of $\rho(r)\sim r^{-2}$, and nine out of twelve cores still have $\alpha<1$ even assuming a constant radial density, suggesting that they are all gravitationally bound, if magnetic field, rotation, or external pressure are not considered. The virial parameters of the three star forming cores in MM1 and MM5 are $\ll1$, therefore these cores are strongly self-gravitating. Even if magnetic field is taken into account, these three cores are still gravitationally bound, consistent with star formation activities associated with them.
\item The virial parameters of the cores in G28.53 and in several hot core sources are all $\sim$0.2--1, even though gas temperatures of hot cores are much higher, suggesting turbulent motions are more important than thermal motions in supporting massive cores against gravity as they collapse. The consistent virial parameters from the lower-mass cores in IRDCs to the more massive hot cores suggest that turbulence is enhanced as cores are accreting, which is essential in maintaining a steady accumulation of dense gas thus forming high-mass stars.
\item The most massive core in MM1 is found at the center of this clump. The energetic outflow traced by CO and a disk-like structure traced by \methanol{} suggest that there might be an outflow-disk system in this core. With a mass of 56~\msol{}, it will form a high-mass star or multiple stellar system.
\item In addition, MM1 harbors another three possible star forming sites, as well as seven cores that are likely quiescent but also massive. Therefore MM1 might be a progenitor of high-mass star cluster, rather than a quiescent clump suggested by previous studies.
\item In contrast, MM5 clump is intermediate in the evolutionary stage and MM7/8 clumps are quiescent. However they have the potential to continue fragment and collapse until star formation starts as in MM1. MM7/8 might be falling into the main part of G28.53 due to its gravitational pull while fragmentation occurs in themselves, which provides a way to form star clusters.

\end{enumerate} 

\acknowledgments

This work was supported by the National Natural Science Foundation of China under Grants 11328301, 11273015, and 11133001, and the National Basic Research Program (973 program No. 2013CB834905). XL acknowledges the support of Smithsonian Predoctoral Fellowship. KW acknowledges support from the ESO fellowship. 
    
\bibliographystyle{apj}

\clearpage

\begin{landscape}
\begin{deluxetable}{ccccccccc}
\tabletypesize{\scriptsize}
\tablecaption{Summary of the observations. \label{tab:obs}}
\tablewidth{0pt}
\tablehead{
\multirow{2}{*}{Telescope} & \multirow{2}{*}{PI} & \multirow{2}{*}{Project ID} & \multirow{2}{*}{Lines} & \multirow{2}{*}{Date} & \multirow{2}{*}{Pointing\tablenotemark{a}} & \multicolumn{3}{c}{Calibrators} \\
\cline{7-9} & & & & & & Bandpass & Flux & Gain 
 }
\startdata
SMA compact 		& Jonathan Swift & 2009A-H004 & \twelveco{}, \thirteenco{}, \ceighteeno{}, \methanol{} & 2009 05 22 & MM1 & 3C273 & Callisto & 1751+096  \\
SMA compact  		& Jonathan Swift & 2009A-H004 & \fmh{}			& 2009 06 07 & MM1		& 3C273 & Callisto & 1751+096  \\
SMA compact 		& Ke Wang & 2010A-S053 & \twelveco{}, \thirteenco{}, \ceighteeno{}, \fmh{}	& 2010 06 20 & MM5, MM7, MM8 & 3C454.3 & Titan, Neptune & 1911$-$201 \\
\\
VLA D			& Ke Wang & AW763 & \ammone{}, (2,2) 			& 2010 01 04 & A3, A4, A5 	& 3C373    & 3C286 & 1851+005 \\
VLA D			& Ke Wang & AW763 & \ammone{}, (2,2)			& 2010 01 08 & A1, A2		& 3C454.3 & 3C48 & 1851+005 \\
VLA D			& Ke Wang & AW776 & \ammthree{}				& 2010 05 09 & MM1		& 3C454.3 & 3C48 & 1851+005 \\
\\
GBT				& Ke Wang & AGBT10A-067 & \ammone{}, (2,2) 	& 2010 02 28 & 8\arcmin{}$\times$7.5\arcmin{} region & 3C353 & 3C48 & 1852+0055(ptg) \\
\\
VLA C			& Ke Wang & 10B-182 & \fmh{}, \methanol{}		& 2010 11 24 & MA1, MA5	& 3C454.3 & 3C48 & 1851+005
\enddata
\tablenotetext{a}{Coordinates of pointing centers: MM1: (18:44:18.0, $-$3:59:23.0); MM5: (18:44:17.0, $-$4:02:04.0); MM7: (18:44:23.7, $-$4:02:09.0); MM8: (18:44:22.5, $-$4:01:47.0); A1: (18:44:24.0, $-$4:02:12.0); A2: (18:44:17.2, $-$4:02:00.0); A3: (18:44:16.0, $-$4:00:42.0); A4: (18:44:18.0, $-$3:59:35.0); A5: (18:44:18.2, $-$3:58:40.0); MA1: (18:44:18.26, $-$3:59:45.50); MA5: (18:44:18.30, $-$4:01:55.00).}
\end{deluxetable}
\clearpage
\end{landscape}

\begin{deluxetable}{cccccc}
\tabletypesize{\scriptsize}
\tablecaption{Image properties. \label{tab:image}}
\tablewidth{0pt}
\tablehead{
\multirow{2}{*}{Image} & Bmaj & Bmin & Bpa & \multirow{2}{*}{Band/Channel Width} &RMS  \\
& (\arcsec) & (\arcsec) & (\arcdeg) & & (mJy\,beam$^{-1}$) 
 }
\startdata
SMA MM1 cont. & 3.2 & 1.7 & 73 & 4~GHz & 1.1 \\
SMA MM5 cont. & 3.2 & 2.8 & 37 & 8~GHz & 0.9 \\
SMA MM7/8 cont. & 3.1 & 2.7 & 53 & 8~GHz & 0.8 \\
SMA MM1 spec. & 3.1 & 1.6 & 72 & 1.1~\kms{} & 60 \\
SMA MM5 spec. & 3.2 & 2.9 & 38 & 1.1~\kms{} & 50 \\
SMA MM7/8 spec. & 3.2 & 2.9 & 39 & 1.1~\kms{} & 50 \\
VLA+GBT \ammone{}, (2,2) & 3.9 & 2.7 & 20 & 0.3~\kms{} & 3.0 \\
VLA \ammone{}, (2,2) & 3.8 & 2.7 & 20 & 0.3~\kms{} & 3.2 \\
VLA \ammthree{} & 3.4 & 2.6 & $-$161 & 0.2~\kms{} & 4.0 \\
VLA \water{} maser & 1.8 & 1.0 & 34 & 0.8~\kms{} & 1.1
\enddata
\end{deluxetable}

\begin{deluxetable}{cr|lr|lr|lr|lr|lr|l}
\tabletypesize{\scriptsize}
\tablecaption{Fitting results of two velocity components of \amm{} spectra in MM1. \label{tab:2comp}}
\tablewidth{0pt}
\tablehead{
\multirow{2}{*}{Core ID} & \multicolumn{2}{c}{$V_\text{lsr}$\tablenotemark{a}} & \multicolumn{2}{c}{$I_\text{(1,1,m)}$\tablenotemark{b}} & \multicolumn{2}{c}{FWHM} & \multicolumn{2}{c}{$\tau_\text{(1,1,m)}$} & \multicolumn{2}{c}{$T_\text{rot}$} & \multicolumn{2}{c}{$N_\text{\amm}$}  \\
& \multicolumn{2}{c}{(\kms)} & \multicolumn{2}{c}{(mJy\,beam$^{-1}$)} & \multicolumn{2}{c}{(\kms)} & \multicolumn{2}{c}{} & \multicolumn{2}{c}{(K)} & \multicolumn{2}{c}{(10$^{16}$\ \cd)}
}
\startdata
MM1-p1 & 85.9 & 87.7 & 35.0\phantom{$\pm$0.0} & 25.0 & 1.08$\pm$0.05 & 1.34$\pm$0.08 & 5.4$\pm$0.5 & 6.4$\pm$0.7 & 16.9$\pm0.9$ & 14.9$\pm$0.9 & 1.0$\pm$0.1 & 1.3$\pm$0.1 \\
MM1-p2 & 85.2 & 88.0 & 19.0$\pm$1.3 & 25.2$\pm$0.9 & 1.30$\pm$0.08 & 1.33$\pm$0.04 & 3.7$\pm$0.5 & 10.0$\pm$1.1 & 15.1$\pm$0.9 & 13.4$\pm$0.4 & 0.8$\pm$0.1 & 1.9$\pm$0.2 \\
MM1-p3 & 85.1 & 87.8 & 23.3$\pm$1.1 & 21.8$\pm$1.3 & 1.65$\pm$0.07 & 1.31$\pm$0.07 & 6.4$\pm$0.6 & 5.0$\pm$0.6 & 13.7$\pm$0.5 & 14.4$\pm$0.7 & 1.5$\pm$0.1 & 1.0$\pm$0.1 \\
MM1-p4 & 85.7 & 87.8 & 28.1$\pm$1.5 & 18.5$\pm$1.3 & 1.24$\pm$0.05 & 1.24$\pm$0.07 & 6.6$\pm$0.7 & 9.4$\pm$1.8 & 15.5$\pm$0.6 & 13.6$\pm$0.8 & 1.3$\pm$0.1 & 1.7$\pm$0.3 \\
MM1-p5 & 85.1 & 87.9 & 17.6$\pm$1.0 & 22.6$\pm$0.9 & 1.63$\pm$0.08 & 1.21$\pm$0.04 & 5.0$\pm$0.5 & 10.8$\pm$1.4 & 14.4$\pm$0.7 & 13.6$\pm$0.5 & 1.2$\pm$0.1 & 1.9$\pm$0.2 \\
MM1-p6 & 85.8 & 87.8 & 13.0\phantom{$\pm$0.0} & 19.7$\pm$1.2 & 1.74$\pm$0.18 & 1.26$\pm$0.06 & 3.4$\pm$0.6 & 12.7$\pm$2.8 & 15.3$\pm$1.5 & 13.1$\pm$0.8 & 0.9$\pm$0.1 & 2.2$\pm$0.4
\enddata
\tablenotetext{a}{Velocities of the components were all fixed in the fitting.}
\tablenotetext{b}{Intensities of the two components in MM1-p1 and one component in MM1-p6 were fixed to make sure the fitting converged.}
\end{deluxetable}

\clearpage

\begin{landscape}
\begin{deluxetable}{cccccccccc|cr|rcc}
\tabletypesize{\scriptsize}
\tablecaption{Core properties. \label{tab:cores}}
\tablewidth{0pt}
\tablehead{
\multirow{2}{*}{Core ID} & RA & Dec & $V_\text{lsr}$\tablenotemark{a} & Maj.$\times$Min.\tablenotemark{b} & PA\tablenotemark{b} & Flux\tablenotemark{c} & $T_\text{rot}$ & FWHM & \multicolumn{2}{c}{$M_\text{J}$\tablenotemark{d}} & \multicolumn{2}{c}{$M_\text{virial}$\tablenotemark{e}} & $M_\text{virial, \textrm{B}=1\,mG}$\tablenotemark{f} & $M_\text{core}$\tablenotemark{g} \\
 & (J2000) & (J2000) & (\kms{}) & (\arcsec$\times$\arcsec) & (\arcdeg) & (mJy) & (K) & (\kms{}) & \multicolumn{2}{c}{(\msol{})} & \multicolumn{2}{c}{(\msol{})} & (\msol{}) &  (\msol{})
 }
\startdata
MM1-p1 & 18:44:18.04 & $-$03:59:22.81 & 87.7 & 2.2$\times$1.3 & 103 & 84.0$\pm$3.3 & 14.9$\pm$0.9 & 1.34 & 1.9&35.8  & 5.7 & 9.5 & 9.9 & 56.0$\pm$16.5 \\
MM1-p2 & 18:44:17.76 & $-$03:59:34.31 & 88.0 & 5.8$\times$2.2 & 147 & 46.8$\pm$3.6 & 13.4$\pm$0.4 & 1.33 & 1.9&35.8 & 11.7 & 19.6 & 29.5 & 46.2$\pm$13.7 \\
MM1-p3 & 18:44:17.29 & $-$03:59:22.74 & 87.8 & 2.9$\times$1.1 & 22   & 15.0$\pm$2.1 & 14.4$\pm$0.7 & 1.31 & 1.9&35.8 & 5.8 & 9.6 & 13.5 & 7.3$\pm$2.4 \\
MM1-p4 & 18:44:18.02 & $-$03:59:18.96 & 87.8 & 4.0$\times$1.4 & 124 & 17.6$\pm$0.4 & 13.6$\pm$0.8 & 1.24 & 1.9&35.8 & 6.9 & 11.5 & 18.1 & 13.3$\pm$3.9 \\
MM1-p5 & 18:44:17.60 & $-$03:59:29.85 & 87.9 & 3.8$\times$2.7 & 40   & 21.0$\pm$2.1 & 13.6$\pm$0.5 & 1.21 & 1.9&35.8 & 8.9 & 14.9 & 32.1 & 17.2$\pm$5.2 \\
MM1-p6 & 18:44:18.06 & $-$03:59:36.24 & 87.8 & 3.4$\times$2.0 & 89   & 16.3$\pm$0.2 & 13.1$\pm$0.8 & 1.26 & 1.9&35.8 & 7.8 & 13.0 & 20.9 & 16.4$\pm$4.8 \\

MM5-p1 & 18:44:17.30 & $-$04:02:04.88 & 87.2 & 7.3$\times$0.8 & 177 & 20.7$\pm$2.7 & 13.6$\pm$0.4 & 1.06 & 1.4&21.3 & 5.4 & 9.0 & 11.5 & 38.2$\pm$12.0 \\
MM5-p2 & 18:44:17.01 & $-$04:02:01.54 & 86.9 & 4.4$\times$2.0 & 32   & 13.5$\pm$1.2 & 13.8$\pm$0.4 & 1.30 & 1.4&21.3 & 9.4 & 15.7 & 24.6 & 24.3$\pm$7.3 \\
MM5-p3 & 18:44:17.16 & $-$04:02:10.01 & 87.0 & 5.8$\times$2.8 & 40   & 14.6$\pm$0.1 & 12.5$\pm$0.4 & 1.12 & 1.4&21.3 & 9.7 & 16.2 & 40.6 & 30.4$\pm$8.7 \\

MM7-p1 & 18:44:24.06 & $-$04:02:12.60 & 88.2 & 7.9$\times$2.7 & 144 & 19.6$\pm$4.2 & 13.7$\pm$0.4 & 1.00 & 6.0&58.5 & 9.4 & 15.6 & 130.5 & 11.1$\pm$4.0 \\

MM8-p1 & 18:44:22.18 & $-$04:01:43.30 & 88.1 & $<$8.0$\times$2.7 & \nodata & 18.5$\pm$1.4 & 13.5$\pm$0.5 & 1.05 & 2.9&37.9 & 10.2 & 17.0 & 69.9 & 24.7$\pm$7.3 \\
MM8-p2 & 18:44:21.83 & $-$04:01:38.38 & 88.0 & $<$5.4$\times$2.9 & \nodata & 14.5$\pm$0.8 & 14.6$\pm$0.7 & 1.23 & 2.9&37.9 & 11.5 & 19.1 & 58.9 & 17.3$\pm$5.1
\enddata
\tablenotetext{a}{$V_\text{lsr}$ of the cores were determined from cross-correlations between a model and the \ammone{} spectra. For the cores in MM1 that exhibit two velocity components, the ones at $\sim$87~\kms{} are listed (cf. \autoref{tab:2comp}).}
\tablenotetext{b}{Major and minor FWHMs and position angles of the cores were deconvolved from beam. For MM8-p1 and MM8-p2 that are not resolved, upper limits of major and minor FWHMs are given.}
\tablenotetext{c}{Fluxes are corrected for primary-beam response. Errors listed here are from 2D gaussian fittings, not including 20\% uncertainty in flux calibration of the SMA data.}
\tablenotetext{d}{The two columns list thermal Jeans masses and turbulent Jeans masses of the clumps, respectively.}
\tablenotetext{e}{The two columns list virial masses assuming radial density profile of $\rho\sim r^{-2}$ and constant radial density profile respectively.}
\tablenotetext{f}{The total virial mass assuming constant radial density profile and magnetic field strength $B$=1~mG.}
\tablenotetext{g}{For the cores in MM1, the masses have been scaled to those of the 87~\kms{} component, based on the ratios of \amm{} column densities of the two components listed in \autoref{tab:2comp}.}
\end{deluxetable}
\clearpage
\end{landscape}

\begin{deluxetable}{lcccccc}
\tabletypesize{\scriptsize}
\tablecaption{CO outflow properties. \label{tab:outflows}}
\tablewidth{0pt}
\tablehead{
\multirow{2}{*}{Parameter} & \multicolumn{2}{c}{MM1-p1\tablenotemark{a}} & \multicolumn{2}{c}{MM1-p2\tablenotemark{b}} & \multicolumn{2}{c}{MM5} \\
 & Blue & Red & Blue & Red & Blue & Red
 }
\startdata
Velocity range (\kms{}) & [78,87] & [78,87] & [88,96] & [88,96] & [77,88] & [89,99] \\
Excitation temperature (K) & 14.9 & 14.9 & 13.4 & 13.4 & 13.6 & 13.6 \\
Total mass (\msol{}) & 0.84 & 0.63 & 0.71 & 1.08 & 0.16 & 0.24 \\
Momentum (\msol{}\,\kms{}) & 3.76 & 2.66 & 3.57 & 6.06 & 0.58 & 1.66 \\
Energy (\msol\,km$^2$\,s$^{-2}$) & 10.44 & 6.88 & 9.81 & 18.99 & 1.54 & 6.55 \\
Projected lobe length (pc) & 0.52 & 0.92 & 0.60 & 0.71 & 0.31 & 0.68 \\
Dynamic age (10$^4$ yr) & 5.44 & 9.52 & 7.46 & 8.76 & 3.10 & 5.50 \\
Outflow rate (10$^{-5}$ \msol{}\,yr$^{-1}$) & 1.54 & 0.66 & 0.96 & 1.23 & 0.53 & 0.43
\enddata
\tablenotetext{a}{Red and blue lobes are both blue-shifted with respect to the core.}
\tablenotetext{b}{Red and blue lobes are both red-shifted with respect to the core.}
\end{deluxetable}

\begin{deluxetable}{cccccc}
\tabletypesize{\scriptsize}
\tablecaption{\water{} maser properties. \label{tab:masers}}
\tablewidth{0pt}
\tablehead{
\multirow{2}{*}{Maser ID} & \multirow{2}{*}{RA} & \multirow{2}{*}{Dec} & Peak flux & Peak velocity & Velocity range \\
 & & & (Jy) & (\kms{}) & (\kms{}) 
 }
\startdata
W1 & 18:44:16.76 & $-$03:59:17.00 & 6.47 & 85.0 & 84.4--85.6 \\
W2 & 18:44:17.23 & $-$03:58:22.50 & 1.03 & 85.3 & 83.4--86.7 \\
W3 & 18:44:18.23 & $-$04:00:01.50 & 0.24, 0.59, 0.18 & 83.4, 90.1, 97.7 & 81.7--99.4 \\
W4 & 18:44:19.63 & $-$03:59:32.00 & 0.74 & 85.9 & 84.2--87.6
\enddata
\end{deluxetable}

\end{document}